\documentclass[a4,11pt]{article}

\usepackage{amsmath,amssymb,color,graphics,amscd,amsfonts,epsf}
\usepackage{verbatim}
\newcommand{\Slash}[1]{\ooalign{\hfil/\hfil\crcr$#1$}}
\def\coeff#1#2{{\textstyle {\frac {#1}{#2}}}}
\def\half{\coeff 12}
\def\R{{\mathbb R}}
\def\Z{{\mathbb Z}}
\allowdisplaybreaks[4]
\def\tr{{\rm tr}}
\date{}
\begin{document}

\begin{flushright} 
 KUNS-2270\\
WIS/06/10-MAY-DPPA\\
SLAC-PUB-14152\\
\end{flushright} 

\vspace{0.1cm}

\begin{center}
  {\LARGE
  Large-$N$  reduction  in QCD-like theories \\with massive adjoint fermions
   
  }
\end{center}

\vspace{0.2cm}

\begin{center}
         Tatsuo A{\sc zeyanagi}$^{a}$\footnote
         {
E-mail address : aze@gauge.scphys.kyoto-u.ac.jp},
         Masanori H{\sc anada}$^{b}$\footnote
         {
E-mail address : masanori.hanada@weizmann.ac.il}, 
Mithat \"{U}{\sc nsal}$^{b,c}$\footnote
         {
E-mail address : unsal@slac.stanford.edu} 
and 
         Ran Y{\sc acoby}$^{b}$\footnote
         {
E-mail address : ran.yacoby@weizmann.ac.il}

\vspace{0.3cm}

 $^a$           
{\it Department of Physics, Kyoto University,\\
Kyoto 606-8502, Japan}\\

${}^{b}$
{\it Department of Particle Physics and Astrophysics,\\
 Weizmann Institute of Science,\\
     Rehovot 76100, Israel }\\
 
$^c$ { \it SLAC and Physics Department, 
Stanford University, \\
Stanford, California 94025/94305, USA}\\

\end{center}

\vspace{1.5cm}

\begin{center}
  {\bf abstract}
\end{center}
Large-$N$ QCD with heavy  adjoint fermions emulates pure Yang-Mills theory at long distances. 
We study this theory on a four- and three-torus, 
and analytically argue the  existence of a large-small volume  equivalence. 
For any finite mass, 
the center-symmetry unbroken phase
 exists at sufficiently small volume  and this phase   can be used to study the large volume limit  through the Eguchi-Kawai equivalence. 
 A finite-temperature version of 
volume independence implies that thermodynamics  
on $\R^3 \times S^1$  can be studied  via  a unitary matrix 
quantum mechanics on $S^1$, by varying the temperature.    
To confirm this  nonperturbatively, we numerically study both  zero- and 
one-dimensional theories  by using Monte-Carlo simulation.  
The order of finite-$N$ corrections turns out to be $1/N$. 
We  introduce various  twisted versions of the reduced QCD which systematically suppress 
 finite-$N$ corrections.
Using a twisted model, we observe the confinement/deconfinement transition on a 
$1^3 \times 2$ lattice.    The result  agrees with  large volume simulations  of Yang-Mills theory. 
We also comment that the  twisted model can serve as a nonperturbative formulation 
of the  noncommutative Yang-Mills theory.

\newpage
\newpage

\setcounter{footnote}{0}

\tableofcontents

\newpage
\section{Introduction}
\hspace{0.51cm}
Recently Yang-Mills (YM) theory with adjoint fermions, QCD(Adj),  
has attracted much interest.  The main impetus behind this is a network of exact large-$N$ equivalences. 
The starting point is the 
large-$N$ orientifold equivalence,  
which states that the bosonic subsector of this theory is equivalent to the charge-conjugation 
even subsector of QCD  with fermions in antisymmetric representation [QCD(AS)] \cite{ASV03},  provided  symmetries defining the neutral subsectors are  not spontaneously broken 
\cite{Kovtun:2004bz}.  QCD(AS) reduces to the ordinary QCD with fundamental quarks when $N=3$, and is a natural large-$N$ generalization thereof  \cite{CR79}.\footnote{The phenomenology of 
QCD(AS) is examined in \cite{Armoni:2009zq,Cherman:2009fh, HoyosBadajoz:2009hb}.}
The second important link is an orbifold equivalence:  when quarks are massless or light with respect to strong scale $\Lambda_{\rm QCD}$, 
QCD(Adj) compactified on  a   Euclidean four-torus exhibits volume independence,  thanks to its unbroken  prerequisite  (center  and translational)  symmetries
 at any radius  \cite{KUY07}. 
 Thus, 
 through the  Eguchi-Kawai (EK) equivalence \cite{EK82},  one can study  large-$N$ QCD on 
 $\mathbb R^4$ by using  a unitary matrix model on a single-site lattice. 
 QCD(Adj) also  provides new insights into gauge dynamics, especially on  small $S^1 \times \mathbb R^3$. This theory 
  exhibits new nonperturbative phenomena, most strikingly   confinement due to magnetic bions, a new  class of non-self-dual topological excitations \cite{Unsal:2007vu, Unsal:2007jx}, distinct from monopoles and instantons.

The statement of the volume independence \cite{KUY07, EK82, Yaffe:1981vf, 
NN03} is as follows. 
Consider   SU(N) gauge theories on ${\mathbb R^4}$, with or without fermions, toroidally compactified on a four manifold ${\mathbb R^{4-d} \times T^d}$.  
For simplicity, we assume the matter fields  are in adjoint representation and, hence, the theory has a  global $({\mathbb Z}_N)^d$ center symmetry, described most easily as gauge rotations aperiodic 
up to an element of the center group.  
   The order parameters of this symmetry are  Wilson lines  wrapping distinct toroidal cycles. 
The observables singlet under the center transformation constitute the neutral sector. The volume independence states that dynamics of the neutral sector  observables
  is independent of the   size of the torus provided the center symmetry  and translational invariance are not spontaneously broken. Among such observables are 
nonperturbative mass spectrum, free energy densities, and deconfinement transition temperature,  just to count a few.  This is clearly an extraordinarily well justified  reason to study aspects of the small volume, large-$N$ gauge theories.

In fact, center symmetry is spontaneously broken in most  examples. 
In the original Eguchi-Kawai model \cite{EK82}, which is a one-point reduction of 
Wilson's bosonic lattice gauge theory, the breakdown can be shown by one-loop calculation around 
a diagonal background \cite{BHN82}. The quenched \cite{BHN82,Parisi82}
and twisted \cite{GAO82} modifications of the Eguchi-Kawai model
were proposed to  preserve the symmetry,  but after two decades, it is now understood that 
both modifications fail  nontrivially due to nonperturbative effects
 \cite{BS08}, \cite{BNSV06,TV06,AHHI07}.\footnote{The failure of these modifications can be cured  by introducing supersymmetry \cite{AHHI07,AHH08}. Reference \cite{IIST08} proposed  a concrete way to construct  4d ${\cal N}=4$ super Yang-Mills theory  by using the Eguchi-Kawai equivalence, which preserves  16 supersymmetries. 
 Reference \cite{Poppitz:2010bt} studied the Eguchi-Kawai reduction in the strong coupling domain of ${\cal N}=4$ super Yang-Mils   by using AdS/CFT and D-branes. 
} \footnote{Recently, a new limiting procedure for the twisted Eguchi-Kawai model, which aims to prevent center breaking,  has been proposed   \cite{GAO10}. }
Similarly, in any gauge theory compactified thermally on $\mathbb R^3 \times S^1$, or on a torus with at least one thermal boundary condition, center symmetry  breaks spontaneously in the high temperature deconfined phase, and  volume independence is only valid in the low temperature confined phase,  above a  critical volume \cite{NN03}.  

Kovtun, one of us   (M.\"U.),  and Yaffe, motivated by  the  quantitative differences between thermal and  circle (nonthermal) compactifications,  showed that 
if light or massless adjoint fermions endowed with periodic boundary conditions   are added  to Yang-Mills theory,  then 
the center symmetry is stabilized at   small volume dynamically \cite{KUY07}.   
(Also see the discussion in Refs.\cite{Bedaque:2009md, Bringoltz09, PU09}.)

For heavy  fermions,  the infrared physics of  QCD(Adj)  on $\R^4$ emulates the bosonic Yang-Mills theory.  Since  heavy  fermions are also capable of restoring center symmetry at sufficiently small volume, 
this may  provide an opportunity for a working Eguchi-Kawai reduction for  
an ``almost" Yang-Mills theory.   However, this is not straightforward. 
When  one dimension is compactified, QCD(Adj) with massive fermions  on small $S^1 \times \mathbb R^3 $ exhibits an intricate phase structure. This   can be deduced from a 
one-loop effective action of the Wilson line \cite{ Bringoltz09, MO09}, simulations on an asymmetric torus mimicking $S^1 \times \mathbb R^3 $ \cite{Cossu:2009sq}, and studies on $S^1 \times S^3$  \cite{HM09} which also  mimics $S^1 \times \R^3$  
due to topological reasons, as explained in \cite{Unsal:2007fb}.
In all these cases, the ${\mathbb Z}_N$ symmetry along  $S^1$  is intact 
at large radius, and as one decreases 
the radius, it breaks down completely at some critical point and then  gradually restores 
to various   subgroups of the center symmetry. 
${\mathbb Z}_N$ symmetry is restored fully only at  $mLN \sim {\rm few}$, where $L$ is the compactification radius and $m$ is the fermion mass in continuum. This is troubling because 
the volume independence (strong coupling, non-Abelian confinement)  domain is $LN \Lambda_{\rm YM} \gg1$  \cite{UY08},   
whereas, for heavy fermions,  $m \gtrsim \Lambda_{\rm YM}$, the first condition implies  
$LN  \Lambda_{\rm YM}  \lesssim{\rm few}$ which  
is a  volume dependent,  weak-coupling Abelian confinement domain  \cite{UY08}.

One might expect a similar pattern for a single-site lattice  model  based on  this   weak-coupling intuition.  However, recent important work of Bringoltz and Sharpe shows that $({\mathbb Z}_N)^4$ remains  intact in a rather generous  ``funnel,"  in the fermion mass, lattice coupling plane, covering the continuum limit of Yang-Mills theory  \cite{BS09} (also, see \cite{HN09}), corresponding  to the limit where the bare fermion mass is larger than cutoff scale. In particular, they observe a 
$({\mathbb Z}_N)^4$ restoration at $m a \sim O(N^0)$, where $m$ is bare lattice mass and $a$ is lattice spacing. Why and how the full center-symmetry restoration takes place at  $m a \sim O(N^0)$  is the main theoretical problem that we wish  to address in this work.

\subsection{Results}
\label{res}
To set the notation,  we first express the action 
of  continuum  QCD(Adj) on a four manifold:
\begin{eqnarray}
S
=
\frac{N}{\lambda_{4d}}\int_0^\beta dt\int d^3x\  
Tr\Big[
\frac{1}{4}F_{\mu\nu}^2
+ 
 \sum_{f=1}^{N_f^D} \bar{\psi}_f(\Slash{D}+m)\psi_f
\Big],  
\label{QCDadj}
\end{eqnarray} 
where $\psi_f$ are Dirac fermions with mass $m$. (Generalization to different values of masses is  straightforward.) 
In general, we will consider this theory on four-torus $T^3 \times S^1$, with sizes  
 $L$ and  $\beta$, respectively.  
If  we impose 
the antiperiodic boundary condition on fermions along the temporal direction, this action 
describes the finite-temperature system and $\beta$ corresponds to the inverse temperature.  
For periodic boundary conditions on fermions in all directions, we consider $\beta=L$, a symmetric four-torus. 

We show that, with   periodic boundary conditions,   
$({\mathbb Z}_N)^4$ symmetry is not broken at sufficiently small-$L$, although 
it can be  broken at some  intermediate volume.  
(More precisely, $({\mathbb Z}_N)^4$ symmetric and  broken 
phases can coexist, while quantum tunneling between them 
 is suppressed in the large-$N$ limit.) 
The argument, which will be quantified  in Sec.\ref{sec:0d pbc},
 is very simple -- although the one-loop effective action suggests 
the existence of the attraction between Wilson line eigenvalues  at small  separation, 
the eigenvalues  spread due to nonperturbative quantum fluctuations \cite{AHHS09}, and the one-loop calculation  can be trusted only at large separations  where it leads to repulsion. 
The estimates of nonperturbative  fluctuations are outside the reach of one-loop perturbation theory and often overwhelm the implications of one-loop analysis. 
Because of nonperturbative effects, we find that the full center restoration takes place at $mL\sim O(N^0)$, which is compatible  with $LN \Lambda_{\rm YM} \gg 1 $ for   $m \gtrsim \Lambda_{\rm YM}$.

Our main results are
\begin{enumerate}
\item {\bf Small-large volume equivalence:}
The  QCD(Adj) with  heavy fermions of mass $m$ on $\R^4$, or  $T^4$  with radii $L \gtrsim \Lambda^{-1}$,   is equivalent to the theory 
 on a small $T^4$, with radii $mL < O(N^0)$,   due to unbroken  $({\mathbb Z}_N)^4$ 
center symmetry associated with the cycles on  $T^4 $. This is a small-large volume equivalence  with an intermediate center-symmetry broken phase, where volume independence is not valid.  This is depicted in the following figure. 
\begin{figure}[htbp]
\begin{center}
\scalebox{1.5}{
\includegraphics{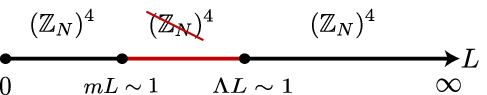}}
\end{center}
\label{s-l}
\end{figure}
\item{\bf Finite-temperature equivalence:}
The theory on $\R^3 \times S^1$ at finite temperature is equivalent 
to the one on $T^3 \times S^1$ provided  the $({\mathbb Z}_N)^3$ 
center symmetry associated with the cycles on  $T^3 $ is not spontaneously broken. 
The  phase transition in the thermodynamic $\R^3 \times S^1$ limit  can be studied by 
using a large-$N$ reduced model on small $T^3 \times  S^1$ by dialing $\beta$.  
This form of equivalence is also useful for Hamiltonian  formulation and extraction of the nonperturbative spectrum of the theory. 

\item{\bf Twisted QCD:}  In compactified QCD(Adj), finite-$N$ corrections  turn out to be order $1/N$,  as opposed to perturbative  expectation  on $\R^4$ \cite{'tHooft:1973jz}, which is order 
$1/N^2$.
There are two plausibly  related  explanations for this behavior, whose footprints can be seen in  
finite-volume perturbation theory. 
In a weak-coupling   center-symmetric background,   the volume is only enhanced by a factor of $N$,  and the effective volume is  $V_{\rm eff} \sim NV$. Finite-$N$ corrections should  scale as finite-volume corrections.     The other is, in compact space, one cannot gauge away zero-momentum modes. Typically, there are order $N$ bosonic and fermionic zero modes, which may generate nonperturbative $1/N$ effects  \cite{BS10}. 
Both problems  can simultaneously be cured  and  $1/N$ corrections can systematically be improved  by using twisted boundary conditions.\footnote{The relation between ``effective" volume and $N$ is discussed in many places, for a review, see \cite{Das:1984nb}. 
The observations  that twisted boundary conditions can be used   {\it i)} to  
 systematically reduce finite-volume corrections is given in  Ref.\cite{Coste:1986cb} and  {\it ii)}  to  lift bosonic and fermionic zero modes is given in Ref.\cite{Witten:1982df}. 
   We give analytic and numerical evidence 
 suggesting that the two are   indeed related.} We refer the latter as TQCD(Adj) as per  \cite{GAO82}.
 \end{enumerate}

Our results have interesting spin-offs for noncommutative theories, phase transition in pure Yang-Mills theory and orientifold equivalence. 
 Adding massive or massless adjoint fermions to the twisted Eguchi-Kawai model (TEK) 
 model cures the global instability of the model \cite{BNSV06,AHHI07,TV06}. Therefore, our formulation can be used to provide a nonperturbative definition of noncommutative bosonic Yang-Mills theory and noncommutative  QCD(Adj). 
By using the reduced matrix model for TQCD(Adj) with  very heavy  adjoint fermions on $1^3 \times 2$ lattices,  
we  observe the confinement/deconfinement transition at bare coupling $b_c=0.32-0.33$.  
Large volume simulations for pure Yang-Mills theory give similar results, 
in accordance with the  finite-temperature version of equivalence
(For example, $SU(8)$ YM theory simulated on $ 10^3 \times 5$ lattices gives, in the extrapolated infinite volume limit,  $b_c \sim 0.34$  \cite{Lucini:2005vg} (see also \cite{Panero:2009tv}). 
The small difference is due to $1/N$ effects,  the existence of heavy fermions in our reduced model and the difference of numbers of sites along the temporal direction). 
We hope that the unitary matrix model can be used to gain  insight 
into the nature of the deconfinement transition of the infinite volume theory. 

Finally, combined with the large-$N$  orientifold equivalence between (T)QCD(Adj) and  QCD(AS),   
thermal properties of the QCD(AS)  can also be studied by using 
the large-$N$ reduced model for TQCD(Adj)  at finite temperature 
\cite{ASV03,  Kovtun:2004bz,Unsal:2007fb}.\footnote{Earlier work on  deconfinement transition in large-$N$ QCD(Adj) and QCD(AS)  in the  weak-coupling limit of small
$S^3 \times S^1$ showed that the critical temperatures agree \cite{Unsal:2007fb}, 
up to $1/N$ corrections.}   
Our results for massive QCD(Adj)  can be regarded as a first step towards this direction of research.

\section{Review of  original Eguchi-Kawai proposal}
\label{EK}
In this  section, we  give a  brief review of   the  Eguchi-Kawai equivalence. Our aim is not to repeat  the original works on the subject;
  rather we wish to point out two important  ingredients which will be useful.  These are   perturbative  and nonperturbative quantum fluctuations in the matrix model. 

The EK model is the dimensional reduction of Wilson's lattice gauge theory action 
down to a  $1^4$  lattice, i.e., a one-site matrix model. The action of the reduced model is 
\begin{eqnarray}
S_{0d}
=
-2b N
Re\ Tr \left(
\sum_{\mu<\nu}
V_\mu V_\nu V_\mu^\dagger V_\nu^\dagger
\right),
\label{OEK}
\end{eqnarray}
where $V_\mu\ (\mu=1,2,3,4)$ are unitary matrices. 
$b$  is the bare inverse 't Hooft coupling constant. To probe continuum physics,  $b$ should be chosen appropriately depending on the lattice spacing $a$.  
This action has a $({\mathbb Z}_{N})^4$ global  center symmetry 
\begin{eqnarray}
V_\mu
\to
e^{2\pi in_{\mu}/N}V_\mu\qquad (n_{\mu} =0,1,\cdots,N-1). 
\end{eqnarray}    
 The reduced   model  (\ref{OEK})  is equivalent to the translationally invariant subsector of  the lattice theory  with an arbitrary number of sites, provided  center  and translational  symmetries are   not broken.  These are the necessary and sufficient symmetry realization conditions for the large-$N$ volume independence.  

\subsection{Perturbative  fluctuations and one-loop potential}
\label{EKpert}
The realization of center symmetry can be determined by integrating out weakly coupled perturbative modes in the background of diagonal, commuting Wilson lines: 
\begin{eqnarray}
 V_\mu={\rm diag}(e^{i\theta_\mu^1},\cdots,e^{i\theta_\mu^N}), \qquad [V_{\mu}, V_{\nu}] =0.
 \label{comsad}
\end{eqnarray}
The resulting one-loop action reads  (we parallel the  discussion in the one-site matrix model and  continuum theory on small  $T^4$) 
\begin{eqnarray}
S_{\rm 1-loop}  =\left\{
 \begin{array}{ll}
2\sum_{a<b}\log\Big[
\frac{4}{a^2}\sum_{\mu=1}^4\sin^2\Big (\frac{\theta_\mu^{ab}}{2}\Big)   \Big], & \qquad  {\rm on \;  1^4- lattice} 
 \cr \cr
2  \sum_{a<b}  \sum_{ \vec k \in \Z^4}
     \;  \log  \left( \sum_{\mu=1}^4  \frac{(2 \pi k_{\mu} + \theta_{\mu}^{ab})^2}{L^{2}} \right), & \qquad   {\rm on \;  T^4-continuum }
\end{array}
\right.
\label{1loop-EK}
\end{eqnarray}
where $ \theta_{\mu}^{ab}=  \theta_{\mu}^{b}- \theta_{\mu}^{b}$. Note that one-site matrix model and $T^4$  have the same symmetry properties, both invariant under $ \theta_{\mu}^{ab} \rightarrow \theta_{\mu}^{ab} + 2\pi$.\footnote{The relation between the continuum one-loop effective action on $T^4$  and  
the one-loop effective action for one-site theory given in Eq.
(\ref{1loop-EK}) is analogous  of the one  between the action of the XY model and its Villain form.}  
It is also convenient to rewrite  (\ref{1loop-EK})  by using the Wilson lines. This can be done by a 
Fourier transformation for the former and by a    Poisson resummation for the latter:
\begin{eqnarray}
    S_{\rm 1-loop}[V_1, \ldots,  V_4 ] =\left\{
 \begin{array}{ll}     
         \sum_{ \vec n \in \Z^4 \setminus \{{\bf  0}\}}   
           P(\vec n)
          \left( \left|\tr\left(V_1^{n_1}    \cdots V_4^{n_4} \right)\right|^2 -N\right),  
&  \qquad  {\rm on \;  1^4- lattice} 
 \cr \cr
    - \frac{1}{\pi^2}  
         \sum_{ \vec n \in \Z^4 \setminus \{{\bf  0}\}}    \frac{1}{ |\vec n|^4}
          \left( \left|\tr\left(V_1^{n_1}    \cdots V_4^{n_4} \right)\right|^2 -N\right),  
& \qquad   {\rm on \;  T^4-continuum }
\end{array}
\right.
\label{1loop-EKFT}
\end{eqnarray}
where
\begin{eqnarray}
P(\vec n)   = - \int \frac{d\alpha}{\alpha}   e^{- 8 \alpha} \;    I_{n_1}(2 \alpha )  \ldots  
 I_{n_4}( 2 \alpha )\,, 
 \label{FC}
\end{eqnarray} 
and   $I_n(2 \alpha  )$ is the  modified Bessel  function of the second kind; 
$I_n(2 \alpha  ) = (2 \pi)^{-1}
   \int_0^{2 \pi} d\theta
 \;\; e^{ 2 \alpha  \cos \left(  \theta \right) }  \; e^{i  \theta  n} $. 
In the one-site model,   Fourier coefficients are slightly more complicated, but the main result is the same as in continuum $T^4$. 
The basic point is   the  negative definiteness of the integral in  (\ref{FC}). Moreover,  at a large-winding number, the integral 
 can be evaluated analytically by localization, and converges to  the continuum $T^4$ result quickly:   
 \begin{eqnarray} 
P(\vec n)   \leq 0, \qquad    \forall \vec n   \in \Z^4 \setminus\{ {\bf  0}\};  \qquad \qquad 
 P(\vec n)  \approx     - \frac{1}{\pi^2}     \frac{1}{ |\vec n|^4}, \qquad { \rm for}  \; 
 |\vec n| \gg 1\, .
 \end{eqnarray}
The one-loop action (\ref{1loop-EK}) has IR singularities whenever two (or more) eigenvalues coincide. This is also manifest in (\ref{1loop-EKFT}); both series are conditionally convergent at a large-winding number, and exhibit logarithmic divergence for coinciding eigenvalues. These issues are discussed  throughly in the Appendix \ref{Poisson}.  
Physically, at these points, there are extra massless degrees of freedom which should not have 
been integrated out. In other words,  the zeroth order assumption that one can expand the fluctuations around commuting saddles  (\ref{comsad})  is not  always    correct.  
As discussed in the Appendix \ref{Poisson}, the IR divergence can be regularized and meaningful results can be extracted from (\ref{1loop-EK}). The result is, of course, well-known;  (\ref{1loop-EK}) generates an eigenvalue attraction, or in   (\ref{1loop-EKFT}) the ``masses" for Wilson lines are all negative and, consequently,  eigenvalues clump.  However, neither implies that  all the eigenvalues are coincident. Because of nonperturbative effects, the eigenvalues spread.  In order to estimate the size of the eigenvalue bunch,  we may study the theory around one of its global minima, $V_{\mu} \approx 1$.

\subsection{Nonperturbative fluctuations and size of eigenvalue bunch}
Parametrizing 
$V_{\mu} \approx e^{ia X_{\mu}}$, we observe that around the minimum of the one-loop potential, all the fluctuations are quartic (as opposed to being quadratic). The action expanded around  one of the minima, the $V_{\mu} \approx 1$ configuration gives   
\begin{eqnarray}
S_{0d}
&=&
\frac{N}{\lambda_{0d}}
\ Tr
\Biggl(-
\frac{1}{4}[X^{\mu},X^{\nu}]^2  \Biggr),
\label{MMYMaction}
\end{eqnarray}
where 
\begin{eqnarray}
\lambda_{0d}=   \left\{
\begin{array}{ll}     
    \frac{1}{2 b a^4} 
          &  \qquad  {\rm on \;  1^4- lattice} 
 \cr \cr
 \frac{\lambda_{4d}(\frac{1}{L})}{L^4}   \sim \frac{1}{\log (\frac{1}{L\Lambda}) L^4}
& \qquad   {\rm on \;  T^4-continuum }
\end{array}
\right.
\label{0tH}
\end{eqnarray}
is the zero-dimensional 't Hooft coupling.   In the lattice, $b$ is, of course, a  parameter that one can choose at will. However, to probe continuum physics,  it needs to scale as 
$ b \sim  \log (1/a \Lambda)  $
by asymptotic freedom.  

 Since the Hermitian matrix model is given in terms of noncompact matrices, it is not {\it  a priori}  guaranteed that the theory  actually  exists quantum mechanically. 
   Reference \cite{Austing:2001bd} shows that     the  
 partition function of the theory defined through (\ref{MMYMaction})  does not exist for SU(2), but exists for SU(N)  for $N\geq 3$.   In the next section, we will indeed see  closely related  Hermitian matrix  models  which do not exist for any $N$.  Despite this, 
the Hermitian matrix model defined through  (\ref{MMYMaction})  is useful. 
It can be employed to determine the   interaction between eigenvalues and to 
estimate the  root-mean-square fluctuations of $X_{\mu}$ matrices. (We also refer to this as the 
size of the eigenvalue bunch.)
 
Scalar eigenvalues in 0d theory are related to the phases $\theta$ of the Wilson lines  
winding on temporal and spatial directions by 
\begin{eqnarray}
L{x}^{\mu}_a = \theta^{\mu}_a
\label{xtheta}
\quad
(a=1,\cdots,N). 
\end{eqnarray}
The 't Hooft coupling $\lambda_{0d}$ has the dimension of $(mass)^4$ and its value 
sets the typical mass scale of the 0d theory (\ref{MMYMaction}).   
In particular, typical fluctuation of the eigenvalues of the dynamical fields is set by this scale.  
Because of the generic noncommutativity of $X_{\mu}$ matrices , the relative positions of the eigenvalues make  sense only when their separation is of order or larger than 
$\lambda_{0d}^{1/4}$ \cite{AHHS09}. 

The one-loop effective action of the matrix model
around the diagonal background can be  calculated  if all eigenvalue pairs are well-separated 
$ |\vec x_{a}-\vec x_{b}| \gg \lambda_{0d}^{1/4} $,  corresponding to the case where all eigenvalues are weakly coupled.  Integrating out massive ``W bosons" (off-diagonal elements)  by using 
the background field gauge yields
\begin{eqnarray}
S_{\rm 1-loop}[  x^{\mu}_{ab}] 
&=&
2\sum_{ a< b}
\log |\vec x_{a}- \vec x_{b} |^2.
\label{1loop,EK3}
\end{eqnarray}
 Clearly, this  is nothing but  Eq.(\ref{1loop-EK}) restricted to its Kaluza-Klein (KK) zero mode. 
 Thus, as in the one-loop potential  (\ref{1loop-EK}) of the  full theory, the Hermitian model as well 
 predicts eigenvalue attraction  and also exhibits the same IR singularity whenever two eigenvalues coincide. 

More importantly, the  Hermitian matrix model   (\ref{MMYMaction})  allows the determination of the  typical size of eigenvalue fluctuations. In  't Hooft's large-$N$ limit, 
the root-mean-square fluctuations $\Delta X$ are
\begin{equation}
\Delta X =  
\sqrt{ \Big\langle \frac{1}{N} \tr (X_{\mu}^2) \Big\rangle } \sim  \lambda_{0d}^{1/4}\,.
\end{equation}
This result has various interesting implications. 

The analysis of continuum $T^4$ and weak-coupling  ($b \rightarrow  \infty$) domain of lattice gauge theory exhibits similar   behavior. Let $L=a$. Then, $\Delta X  \sim  
1/\left[ \log (1/L\Lambda) \right]^{1/4} L  <  1/L$. At asymptotically small $L$, 
the  size of the eigenvalue bunch     is much smaller than the size $2\pi/L$  of the dual torus where eigenvalues are residing. In this domain, clearly, nonperturbative fluctuations cannot overwhelm broken center symmetry.

It is well-known that Eguchi-Kawai reduction holds in the strong coupling domain of lattice gauge theory, for $0\leq b < b_c= 0.19$.  This is a lattice domain unrelated to continuum physics, whereas the above analysis is valid for sufficiently large $b$.  What happens to $\Delta X$ if we  decrease  $b$   to being of order few? Of course, this is unjustified, but  demonstrates
the   trend of the root-mean-square fluctuations.  Naive use of the 
$\Delta X \sim   1/b^{1/4} a$ shows that when  $b^{1/4}$ is order one, the quantum fluctuations may be as large as the size of the dual torus. In this domain,  eigenvalues feel the size of the compact space they live in  and  we expect the strong fluctuations to lead to center restoration.  This is indeed the case.

In  this paper, we will study Yang-Mills theory with massive adjoint fermions  with mass $m$.  When $m=\infty$, the discussion is the same as the original Eguchi-Kawai reduction.  
For $ma \sim {\rm few}$,  the perturbative-loop analysis indicates that only a  finite  subgroup of center symmetry would restore; however, such analysis does not take into account nonperturbative  fluctuations.  We propose that these fluctuations tend to   uniformize the eigenvalue distribution rather quickly, even when $ma \sim {\rm few}$, providing   the resolution of the problem  stated in the introduction.

\section{Eguchi-Kawai equivalence for massive QCD(Adj)}

In this section we  explain why it is natural to expect that the Eguchi-Kawai reduction holds 
for the QCD(Adj) with massive adjoint fermions at sufficiently small volume. 
Our analysis is based on the one-loop potential for diagonal components of fields and estimates 
of nonperturbative  quantum fluctuations. 
We start with the zero temperature case and then generalize to the finite temperature.

\subsection{QCD(Adj) at zero temperature on $T^4$ at small  $L$ } \label{sec:0d pbc}
\hspace{0.51cm}
Let us consider the QCD(Adj) at zero temperature on a symmetric four-torus, with size $L$.   
At sufficiently small $L$, 
gauge coupling at the scale of the compactification is small and 
we  may analytically  compute the one-loop effective action on 
$T^4$. 
  The reason for doing this computation, apart from trying to determine the center-symmetry realization,    is twofold. One is we would like to compare with  the  one-site theory, given in  (\ref{1loop-l}). 
 More importantly,  we give evidence that  some (not all)  implications of  one-loop action   are,  in full theory,  overwhelmed by large  nonperturbative quantum fluctuations,  and therefore, incorrect.

The one-loop  effective action,   in the Wilson line  background  (\ref{comsad}),  induced by gauge and  fermionic fluctuations with periodic boundary conditions  can be written as 
\begin{eqnarray}
S_{\rm 1-loop}[\theta_{\mu}^{ab}] =   \sum_{a<b}  \sum_{k_1, \ldots, k_4}  \left(   2 \;  \log 
\left( \sum_{\mu=1}^4  \frac{(2 \pi k_{\mu} + \theta_{\mu}^{ab})^2}{L^{2}} \right)  - 4N_f^D \;  \log \left( \sum_{\mu=1}^4  \frac{(2 \pi k_{\mu} +  \theta_{\mu}^{ab})^2}{L^{2}}  +m^2 \right) \right), \qquad
\label{smallT4}
\end{eqnarray}
where $\theta_{\mu}^{ab}= \theta_{\mu}^{a}- \theta_{\mu}^{b}$, and $m$ is the fermion mass. 
Similar to  original EK model (\ref{1loop-EK}),  this expression  has IR singularities whenever two (or more) eigenvalues coincide, which we discuss throughly 
in the Appendix \ref{Poisson}.  
 The theory may have different saddles which are  expressed in terms of noncommuting matrices. The classification of  the saddles, using  the techniques of Ref.\cite{Coste:1985mn},  
of QCD(Adj) as a function of mass of the fermion is given in the Appendix \ref{Poisson}.

When we consider  a regime where $|\vec \theta^{ab}| \ll 2\pi$,  we can split 
(\ref{smallT4}) into the zero  and non-zero-momentum contribution. 
 The eigenvalue dynamics   is dominated by the interactions between nearby eigenvalues 
 and the effect of high KK modes is negligible. Therefore, we 
 will use  the truncated Hermitian matrix model  (\ref{MM action}) to gain an understanding of 
 the  typical eigenvalue fluctuations.  Strictly speaking, the truncation can be justified only when 
the $({\mathbb Z}_N)^4$ center symmetry is completely broken  and  consequently  there exists a clear separation of scales between the KK modes and zero modes. 
In a center-symmetric background, this is not the case. However, at large-$N$, the states obtained by quantizing the theory on a center-symmetric vacuum  fill the $[0, 2\pi/L]$ energy range uniformly. If we consider a finite but small  range  $ |\vec \theta^{ab}| \ll 2\pi$, there are still  $O(N^2)$ states (in perturbation theory)  in this  interval. We may therefore use the Hermitian model 
to probe the interaction of nearby eigenvalues, and their fluctuations.  The 
eigenvalue dynamics of the full theory is mimicked rather accurately 
 by the  truncated Hermitian matrix model.

The truncation of   the KK modes  in (\ref{QCDadj}) yields the  zero-dimensional  Hermitian matrix model
\footnote{It should be understood that the  Hermitian model  (\ref{MM action}) is 
used as auxiliary.   In the strict $L=0$ limit,  and for massless fermions,  the partition function  of the Hermitian model  diverges. One-loop potential is unbounded from below as eigenvalue separations tend to infinity.  At finite-$L$, the target space of eigenvalues is bounded,  cannot run-off to infinity and   the partition function of the  finite-$L$ theory exists.    }
\begin{eqnarray}
S_{0d}
&=&
\frac{N}{\lambda_{0d}}
\ Tr\Biggl(
-
\frac{1}{4}[X^{\mu},X^{\nu}]^2 
+
\sum_{f=1}^{N_f^D}
\bar{\psi}_f\left(
\gamma_{\mu}[X^{\mu},\psi_f ]
+
m\psi_f
\right)
\Biggl), 
\label{MM action}
\end{eqnarray}
where  $\lambda_{0d}
=\lambda_{4d}(1/L)/L^4 \sim 1/\log (1/L\Lambda) L^4$ 
is the zero-dimensional 't Hooft coupling, and $\Lambda$ is the strong scale of QCD(Adj). 

The  $\lambda_{0d}^{1/4}$ and $m$  are  the two scales in the Hermitian matrix model (\ref{MM action}).   
In particular, for the $m=\infty$ theory,  which corresponds to the original EK model discussed in 
Sec.\ref{EK}, the root-mean-square fluctuations of matrices are set by this scale 
$\Delta X(m=\infty) \sim   \lambda_{0d}^{1/4}$.   With standard  't Hooft scaling,  
 this  is  $O(N^0)$ in the large-$N$ limit.   
This is a  crucial observation that will be important below. 
Because of the generic noncommutativity of $X_{\mu}$ matrices , the relative positions of the eigenvalues make  sense only when their separation is of order or larger than 
$\lambda_{0d}^{1/4}$ \cite{AHHS09}.

The one-loop effective action of the matrix model
around the diagonal background can be  calculated  if all eigenvalue pairs are well-separated 
$ |\vec x_{a}-\vec x_{b}| \gg \lambda_{0d}^{1/4} $,  corresponding to the case where all eigenvalues are weakly coupled.  Integrating out massive ``W bosons" (off-diagonal elements)  by using 
the background field gauge yields
\begin{eqnarray}
S_{\rm 1-loop}[  x^{\mu}_{ab}] 
&=&
2\sum_{ a< b}
\log |\vec x_{a}- \vec x_{b} |^2
-
4N_f^D\sum_{ a< b}
\log\left(
|\vec x_{a} -\vec x_{b}|^2+m^2
\right). 
\label{1loop,pbc}
\end{eqnarray}
 Note that this is nothing but  Eq.(\ref{smallT4}) restricted to its KK zero mode, as  expected. 
 
 The effect of adjoint fermions, which is most manifest in (\ref{1loop,pbc}), is to generate repulsion among eigenvalues.  At $m=\infty$, since the fermions do not contribute, they have no  impact on $\Delta X$.   
For finite fermions' mass, the effect of fermions makes $\Delta X$ larger than 
that of the $m=\infty$ case. Although we will not prove this statement, it is easy to understand it on physical grounds.  When fermion mass is zero, then the one-loop potential is unbounded from below only at  {\it large} eigenvalue separation; hence $\Delta X(m=0) =\infty$.  
From now on, we assume 
$\Delta X(m) \geq\Delta X(m=\infty) = \lambda_{0d}^{1/4}  $. Somewhat conservatively, we use 
the smaller of the fluctuations in what follows.  

Let us now discuss the realization  of $({\mathbb Z}_N)^4$ symmetry based on the  effective action (\ref{1loop,pbc}) for the Hermitian matrix model.

\subsubsection{$N_f^D=0$ (bosonic)}
\hspace{0.51cm}
The $N_f=0$ or $m=\infty$ limits are same and are discussed  in Sec.\ref{EK}. 
We repeat the main result for convenience. 
The  one-loop action leads to  the mutual  attraction of eigenvalues  at 
large eigenvalue separation  and hence, eigenvalues must  clump. This implies  broken 
center symmetry.  However at small eigenvalue separation,   
one-loop approximation  is not valid; the background of the commuting  Wilson line saddles breaks down.   Because of nonperturbative quantum 
fluctuations,  the eigenvalues do not collapse to a point; rather 
the eigenvalue clump has a finite extent of  the order   $ \Delta X \sim \lambda_{0d}^{1/4}\equiv [\lambda_{4d}(1/L)]^{1/4}/L$. Note that, although the size of the eigenvalue clump is suppressed with respect to $1/L$  at small  $\lambda_{4d}(1/L)$
 and hence center symmetry is broken,  it scales as  $O(N^0)$ in the large-$N$ limit. This will be crucial later. 
\subsubsection{$N_f^D=1/2$ (single Majorana)}
\hspace{0.51cm} \label{singleMaj}
When $N_f^D=1/2$ and $m=0$, the theory is 4d ${\cal N}=1$ pure super Yang-Mills theory.  
In this case,  the one-loop effective action \eqref{1loop,pbc} vanishes and in fact, this is true to all loop orders due to supersymmetry.  Taking nonperturbative fractional instanton effects into account, the center is unbroken on $\R^3\times S^1$ as discussed in  \cite{KUY07}. In supersymmetric theories with supersymmetry preserving boundary conditions, it is expected that there are no phase transitions as the volume is varied \cite{Witten:1982df}.   At large-$N$, the 
absence of phase transitions  transmutes to exact volume independence   \cite{KUY07}  and  unbroken  $({\mathbb Z}_N)^4$ center symmetry.  However, it is also possible to  construct a metastable  center-broken sector  \cite{HK09}, which becomes stable at large-$N$. 

When $m$ is fixed and nonzero, by taking $L \rightarrow 0$,  \eqref{1loop,pbc} is positive, and 
therefore, $({\mathbb Z}_N)^4$ symmetry breaks down.  If $L$ is fixed and $m \rightarrow 0$, a center-symmetry preserving background exists for a finite range of $m$. This aspect is discussed 
throughly  in   Ref.\cite{UY10}.  This noncommutativity of  limits requires care in drawing conclusions about this case.

\subsubsection{$N_f^D\ge 1$: Uniformization of  eigenvalue distribution}
\begin{figure}[htbp]
\begin{tabular}{cc}
\begin{minipage}{0.45\hsize}
\begin{center}
\scalebox{0.2}{
\includegraphics{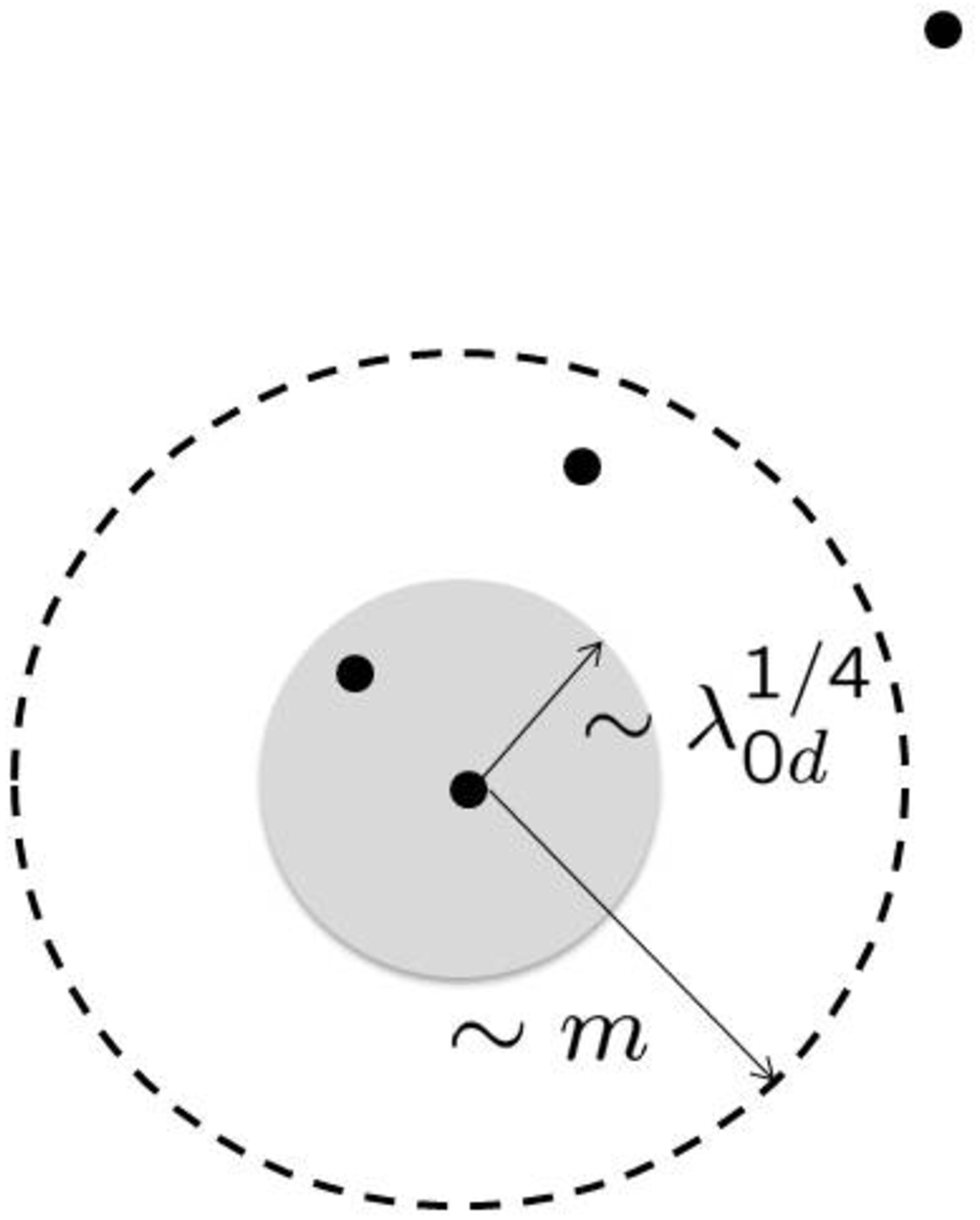}}
\end{center}
\end{minipage}
\begin{minipage}{0.1\hsize}
\end{minipage}
\begin{minipage}{0.45\hsize}
\begin{center}
\scalebox{0.2}{
\includegraphics{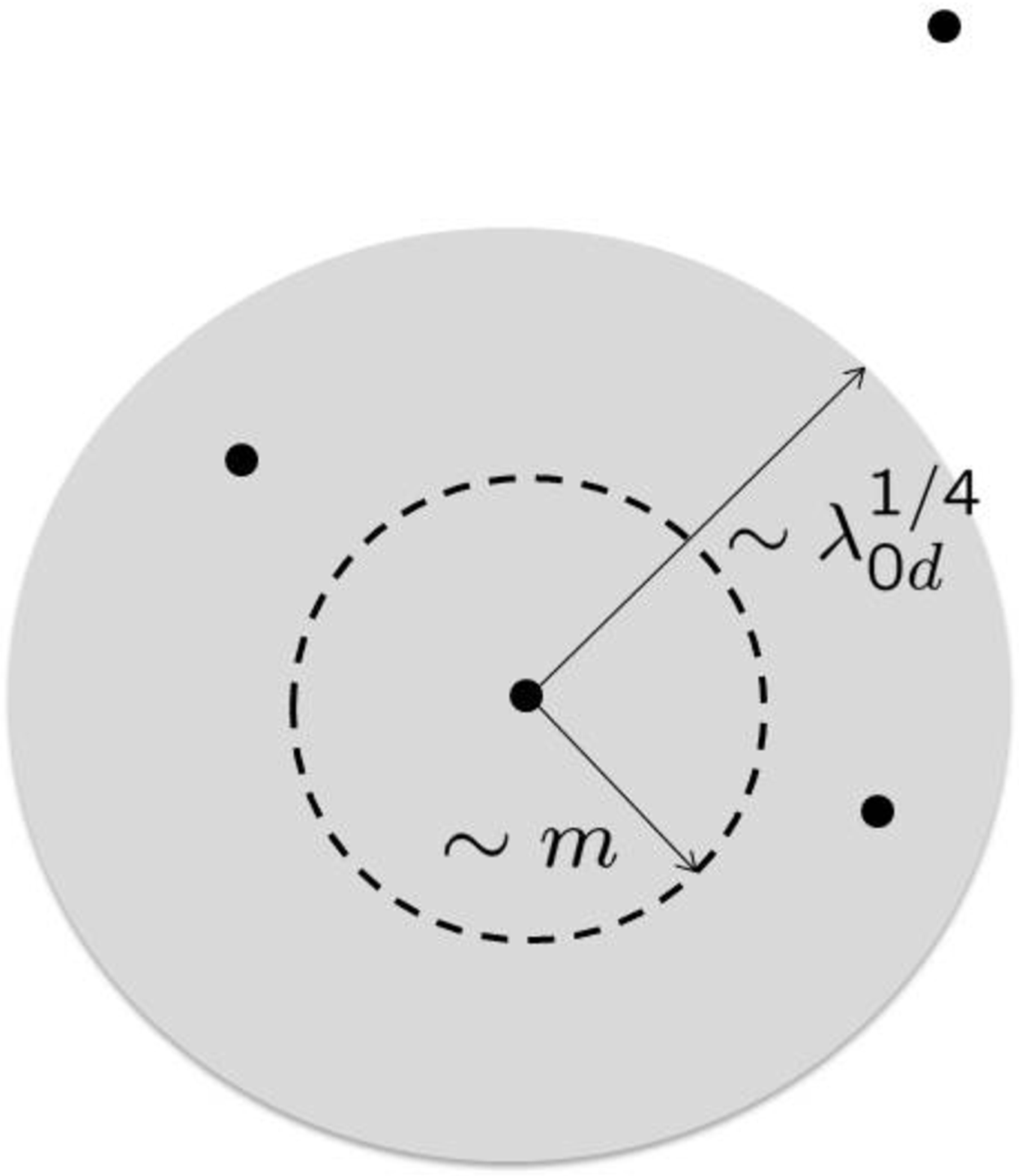}}
\end{center}
\end{minipage}
\end{tabular}
\caption{The scales in the problem. Left panel: $m\gtrsim \lambda_{0d}^{1/4}$ and mass is important.  Right panel: $m\lesssim \lambda_{0d}^{1/4}$ and mass is negligible with respect to 
quantum fluctuations. We can always realize the latter case by taking the size of the four-torus sufficiently small, but still $O(N^0)$. See the text for explanations.  
}\label{region1and2}
\end{figure}

\hspace{0.51cm} \label{mainres}
The effective action \eqref{1loop,pbc} predicts attraction at short distance 
$|\Delta x|\lesssim m$, and repulsion at long distance $|\Delta x|\gtrsim m$. 
Then one may naively conclude that all eigenvalues clump and 
$({\mathbb Z}_N)^4$ symmetry is broken. 
However, one should notice that  this effective action is valid only at 
$|\Delta x|\gtrsim \lambda_{0d}^{1/4}$.   
When $\lambda_{0d}^{1/4}\lesssim m$
attractive force emerges at $\lambda_{0d}^{1/4}\lesssim |\Delta x|\lesssim m$ 
and hence $({\mathbb Z}_N)^4$ is broken (left of Fig.\ref{region1and2}).  
When $\lambda_{0d}^{1/4}\gtrsim m$,  
however,  
it predicts only repulsion;  
eigenvalue fluctuation is of order $\lambda_{0d}^{1/4}$ 
and hence they cannot clump to small region 
where \eqref{1loop,pbc} predicts attraction (right of Fig.\ref{region1and2}). 

At sufficiently small, but $O(N^0)$ compactification radii  $L$,  we can always guarantee that nonperturbative quantum fluctuations 
overwhelm fermion mass, i.e., 
$\lambda_{0d}^{1/4}\equiv [\lambda_{4d}(1/L)]^{1/4}/L \gtrsim  m$. In this case, fermion mass is negligible and for the purpose of center-symmetry realization, the theory cannot be distinguished from the massless theory, for which center symmetry is unbroken. At such  values of $L$,  since the target space of eigenvalues is compact four-torus 
${\widetilde  T^4}$ with size $\frac{1}{L}$,   the repulsion implies that the eigenvalues will uniformly distribute over   ${\widetilde  T^4}$.

 This is the sense in which lower dimensional nonperturbative quantum fluctuations help restoration of center symmetry at 
$mL \sim {\rm few}$, as opposed to quantum field theory on $\R^3 \times S^1$ where full center restoration  requires $mLN \sim {\rm few}$.

As we will see in the Appendix~\ref{Poisson}, the $({\mathbb Z}_N)^4$ broken phase which consists of 
$k$ bunches of eigenvalues may exist when $(\lambda_{0d}/k)^{1/4}\ll m$. 
However, the radius of each bunch is of order $(\lambda_{0d}/k)^{1/4}$, 
which is smaller than the eigenvalue fluctuation in the $({\mathbb Z}_N)^4$ symmetric phase 
$(\lambda_{0d})^{1/4}$, and hence even if 
the $({\mathbb Z}_N)^4$ broken phase exists, tunneling to such a state is highly suppressed.  

The above argument explains the reason for the working of Eguchi-Kawai reduction 
in Refs.\cite{BS09, HN09} for heavy fermions. 
In \cite{BS09}, initial configuration for simulation corresponds to 
$X^\mu=0$ (single bunch). It is unstable and collapses to $({\mathbb Z}_N)^4$ symmetric phase.

\subsection{QCD(Adj) at finite temperature on asymmetric $T^3 \times S^1$}
\hspace{0.51cm}
We can generalize the argument of  the previous subsection 
to  the finite-temperature QCD(Adj) on an asymmetric four-torus $T^3 \times S^1$.   
Circumferences of temporal and three spatial circles are taken to be 
$\beta$ and $L$.  On fermions, we impose antiperiodic 
and periodic boundary conditions along temporal and spatial circles, respectively.  

Completely analogous to the discussion of  Sec.\ref{sec:0d pbc}, 
at small $L$, the  eigenvalue dynamics of the full theory is mimicked rather accurately 
 by the  truncated Hermitian matrix quantum mechanics with action:\footnote{We repeat the same 
 cautionary note as in  Section \ref{sec:0d pbc}. The classical theory  written in Eq.(\ref{MQM action}) does not exist as a quantum theory.  In the strict $L=0$ limit,  and for massless fermions 
 (or fermions with a finite mass), the theory does not have a ground state.     In this paper, 
we study finite-$L$, $ N\rightarrow \infty$ limit of  (\ref{QCDadj}) which is a well-defined quantum theory. The Hermitian  model (\ref{MQM action})    will be  useful as auxiliary to infer  some general lessons about the finite-$L$ case.}  
\begin{eqnarray}
S_{1d}
&=&
\frac{N}{\lambda_{1d}}\int_0^{\beta}dt
\ Tr\Biggl(
\frac{1}{2}(D_tX^i)^2
-
\frac{1}{4}[X^i,X^j]^2 
+
\sum_{a=1}^{N_f^D}
\bar{\psi}_a\left(
\gamma^0D_t\psi_a
+
\gamma_i[X^i,\psi_a]
+
m\psi_a
\right)
\Biggl), 
\label{MQM action}
\end{eqnarray}
where 
\begin{eqnarray}
\lambda_{1d}=\frac{\lambda_{4d}}{L^3}. 
\end{eqnarray}
The 't Hooft coupling $\lambda_{1d}$ has the dimension of $(mass)^3$ and its value 
sets the typical mass scale of the theory.  
In particular, typical fluctuation of the eigenvalues of the dynamical fields is given by this scale.   
The mapping between the scalar eigenvalues and the phases 
 $\theta$ of the Wilson loops is the  same as in   (\ref{xtheta}), but now only running over spatial
  directions.  
Since the $S^1$ direction is compact,  unlike its decompactification limit 
$S^1 \rightarrow \mathbb R$,  the Wilson line along that direction cannot be gauged away.  
We parametrize $W_t=   diag (e^{i\alpha_1 \beta},\cdots,e^{i\alpha_N \beta})$, or  $A_{t} = -i \beta^{-1} \log W_t$.

The one-loop effective action of the matrix quantum mechanics 
around the static diagonal background in the background field gauge 
can be computed in the weak-coupling domain,
 \begin{equation}
\frac{\lambda_{1d}^{1/3} } {| \vec x_{a}-\vec x_{b}|} \ll1\; \qquad  {\rm  or}  \qquad  \lambda_{1d}^{1/3}  \beta \ll 1 \, , 
\label{dov}
 \end{equation}
resulting in
\begin{eqnarray}
S_{\rm 1d,1-loop}(\alpha,\vec{x})
&=&
2\sum_{a<b}\sum_{n=-\infty}^\infty
\log\left(
\left(
\frac{2\pi n}{\beta}+(\alpha_a-\alpha_b)
\right)^2
+
|\vec{x}_a-\vec{x}_b|^2
\right)
\nonumber\\
& &
-
4N_f^D\sum_{a<b}\sum_{n=-\infty}^\infty
\log\left(
\left(
\frac{2\pi (n+1/2)}{\beta}+(\alpha_a-\alpha_b)
\right)^2
+
|\vec{x}_a-\vec{x}_b|^2
+
m^2
\right), \nonumber\\
\label{MQM1}
\end{eqnarray}
where $\alpha_a$ and $\vec{x}_a$ represent diagonal components of 
the gauge field $A_t$ and three scalars $X_i$, respectively. 
By subtracting a constant factor,  (\ref{MQM1})
can be rewritten as \cite{CW07} (for reference, we quote both periodic and antiperiodic 
boundary conditions along $\beta$ circle) 
\begin{eqnarray}
S_{\rm 1d,1-loop}^{\pm}
&=&
2\sum_{a<b}
\log\left(
\cosh\left(\beta|\vec{x}_a-\vec{x}_b|\right)
-
\cos(\beta(\alpha_a-\alpha_b))
\right)
\nonumber\\
& &
-
4N_f^D\sum_{a<b}
\log\left(
\cosh\left(\beta\sqrt{|\vec{x}_a-\vec{x}_b|^2+m^2}\right)
\mp
\cos(\beta(\alpha_a-\alpha_b))
\right). 
\label{1loop_MQM}
\end{eqnarray}
From the  expression of $S_{\rm 1d,1loop}^{-}$, one might naively conclude that 
eigenvalues $\vec{x}_a$ and  $\alpha_a$
coincide,  because the one-loop action is negative infinity at that point. 
However, the analysis is valid only when condition \eqref{dov} is satisfied;
thus drawing reliable conclusions requires care.

The $\beta \rightarrow  \infty$ limit of  (\ref{1loop_MQM}) is both intuitive and insightful.  Since it is the limit of arbitrarily low temperatures, the fermionic boundary conditions should not matter, and indeed, this is transparent from (\ref{1loop_MQM}); the hyperbolic cosine  increases unboundedly, and 
the trigonometric cosine  is bounded, hence the one-loop expression is dominated by the former.\footnote{Also note that at   $\beta \rightarrow  \infty$ limit, $A_t$ can be gauged away. Hence, it should not appear in one-loop potential.} 
The ground-state energy of the system at one-loop order (or one-loop potential) can be deduced from the limit   
$\lim_{\beta \rightarrow \infty} S_{\rm 1d, loop}/\beta \equiv E_{\rm 1d, loop}[|\vec x_{ab}|]$ and the result is
\begin{eqnarray}
&&E_{\rm 1d, 1-loop}[|\vec x_{ab}|] = 2 \sum_{a <b}  |\vec x_a-\vec x_b|  -4N_f^D \sum_{a <b}  \sqrt 
{|\vec x_a -\vec x_b|^2 + m^2},  \cr \cr
&& \lim_{m \rightarrow 0}  E_{\rm 1d, 1-loop}[|\vec x_{ab}|] =  (2-4N_f^D) \sum_{a <b}  |\vec x_a-\vec x_b|. 
\label{energy}
\end{eqnarray} 
This is an   intuitive and simple  result\footnote{ We could have guessed this result by physical reasoning as follows: In the  weak-coupling domain where 
$\widetilde \lambda_{1d}= \lambda_{1d}/|\vec x_{ab}|^3 \ll1$ is small and one-loop analysis is reliable, 
the Lagrangian  (\ref{MQM action}) reduces to a collection of bosonic and fermionic harmonic oscillators:
\begin{equation}
 L \approx  \half |\partial_t \vec X^{ab}|^2  + \half   |\vec x_a-\vec x_b|^2 |\vec X^{ab}|^2 + {\rm fermionic \; oscillators}\,. 
 \end{equation}
 The eigenvalue differences of background $\vec X$ matrices are identified with the frequencies  of harmonic oscillators,   $\omega^{b}_{ab} = |\vec x_a-\vec x_b|$ and 
 $\omega_{ ab}^{f}=   \sqrt {   |\vec x_a-\vec x_b|^2 + m^2}  $. In the chiral limit,  $\omega_{ ab}^{f} = \omega^{b}_{ab} $. 
   In gauge quantum mechanics, there are  $2 \sum_{a < b}$ many massive 
bosonic fluctuations 
and $2 \times 2N_f^D \sum_{a < b}$  many massive fermionic fluctuations. 
Hence, 
the background dependence of the ground-state energy of the system is 
  \begin{equation}
E= \sum_{\rm bosons}  \half \omega^{b} -  \sum_{\rm fermions}  
 \half \omega^f  \;  = \; E_{\rm 1d, 1-loop}[|\vec x_{ab}|]\,, 
\label{SHO} 
\end{equation}
which is given in (\ref{energy}). 
} and various remarks are in order  regarding the chiral $(m=0)$ limit  on $\R$. 
\begin{enumerate}
\item  For $N_f^D > 1/2   $,  $E_{\rm 1d, 1-loop}[|\vec x_{ab}|]$ is unbounded from below and  eigenvalues mutually repel each other.  The classical minima corresponding to the space of commuting triples $[X_i, X_j]=0$ are unstable against perturbative quantum fluctuations.  This means, at  the $L=0$ limit, the 
Hermitian matrix quantum mechanics (\ref{MQM action})  does not have a ground state.  
 \item At finite  $L$,    since the target space of eigenvalues is compact three-torus 
${\widetilde  T^3}$ with size $1/L$,   the repulsion implies that  eigenvalues will uniformly distribute over   ${\widetilde  T^3}$. This implies unbroken center symmetry 
in the $N=\infty$ limit and the theory obeys volume independence.

 \item  For $0\leq N_f^D < 1/2$,  $E_{\rm 1d, 1-loop}[|\vec x_{ab}|]$ is bounded from below.   
The minimum is at $|\vec x_{ab}|=0$. However, in this domain, one-loop analysis 
is not reliable, and there are  quantum fluctuations of  order $\lambda_{1d}^{1/3}$. 
At $N_f^D=1/2$, $m=0$, the ground-state energy is zero to all orders in perturbation theory  due to supersymmetry. 

\end{enumerate}

In the following, we study in more  detail the phase structure of the four-dimensional theory 
by using  (\ref{1loop_MQM}) and estimates of the nonperturbative quantum fluctuations. 
We first explain $N_f^D=0$ (bosonic) and $N_f^D=1/2$ (or one Majorana fermion) cases.  
We then study the case 
with $N_f^D\ge 1$, our main interest in this paper.   
\subsubsection{$N_f^D=0$ (bosonic)}
\hspace{0.51cm}\label{subsec:bosonic MQM}
The  one-loop action  (\ref{1loop_MQM})  generates  an  attraction between eigenvalues 
 at long distance  and  the center symmetry is broken.  More precisely, there are $N^3$ saddles related to each other by center conjugations.  The tunneling between these saddles is suppressed in the large-$N$ limit and the theory is in a center-broken phase. 
   However the eigenvalues 
do not collapse to a single point due to the nonperturbative  quantum fluctuations; 
the clump has a finite size of order $\lambda_{1d}^{1/3} \equiv  [\lambda_{4d}(1/L)]^{1/3}/L$  
which is $O(N^0)$ in the large-$N$ limit.
\subsubsection{$N_f^D=1/2$ (single Majorana)}
\hspace{0.51cm}\label{subsec:MQM_SUSY}
The supersymmetric theory ($N_f^D=1/2$ and $m=0$) has been studied extensively, 
because its maximally supersymmetric cousin has a dual description as the D0-brane system. 
In this case the one-loop potential falls off exponentially, 
\begin{eqnarray}
S_{\rm 1d,1-loop}^{-}
\sim
\sum_{a<b}\exp\left(
-\beta|\vec{x}_a-\vec{x}_b|
\right)\;  , 
\end{eqnarray} 
and the  ground-state energy is $
E_{\rm 1d, 1-loop}[|\vec x_{ab}|] = \lim_{\beta \rightarrow \infty} S_{\rm 1d, 1-loop}^{-}/\beta  \rightarrow 0$. 
This follows from the cancellation of the attractions and repulsions  between eigenvalues 
due to supersymmetry. Consequently, there exists a space of flat directions. Therefore, once eigenvalues are well separated, 
they will propagate freely like the gas of D0-branes to all order in perturbation theory.

In order to study nonperturbative aspects, it is more useful in this case to recall that the theory is formulated on $T^3 \times \R$, and to discuss the system in Hamiltonian formulation. 
As asserted above, to all orders in perturbation theory, the theory has  a moduli space, and all possible realizations of center symmetry are possible. However, non-perturbatively, this is not the case.  Quantization of the zero-momentum bosonic  modes  gives rise to a discrete spectrum and a   gap. 

Witten studied the  gauge quantum mechanics for finite-$N$  within Born-Oppenheimer  approximation and showed that the bosonic  ground-state wave function  (ignoring  fermionic zero modes which are not crucial in what follows) is constant 
\begin{equation}
\Psi_0 ( \vec \theta_1, \ldots, \vec \theta_{N-1}) =  1 
\end{equation}
and the excited states have an energy gap \cite{Witten:1982df}.
The center symmetry and volume independence was not discussed in  Ref.\cite{Witten:1982df}; however, its results have natural implications which apply to our discussion. 
In particular, we can introduce  an eigenvalue distribution function $\rho(\vec \theta)$ measuring the density of eigenvalues.   The density is everywhere non-negative and obeys
$\int d^3 \vec \theta  \rho( \vec \theta) =1$.
Since the ground-state wave function  spreads uniformly over the perturbative flat directions, 
\begin{equation}
\rho  ( \vec \theta) =   \frac{1}{(2\pi)^3}\,, 
\end{equation}
and the center is unbroken. Unlike the discussion in Sec.\ref{subsec:bosonic MQM}, it is also 
unbroken at  large-$N$.   Nonperturbatively, we have a unique saddle singlet under center conjugations, and this is the main difference with respect to purely bosonic theory which has $N^3$ saddles.  

When the fermion is massive,  there is a subtlety due to order of limits of small mass vs. small-$L$ analogous to the discussion  in Sec. \ref{singleMaj}, with similar conclusions.  

Finally,   at large-$N$ ,
this theory has a metastable  bound state of eigenvalues \cite{HK09} (with diverging lifetime  as $N \rightarrow \infty$)  which is analogous to the one in maximally supersymmetric 
theory \cite{AHNT07} corresponding to the black zero-brane 
in type IIA supergravity.

\subsubsection{$N_f^D\ge 1$}
\hspace{0.51cm}\label{subsec:MQM_Nf > = 1}
First consider the   case with  $m=0$. 
From the one-loop effective action \eqref{1loop_MQM}, it is apparent that 
repulsive force coming from fermions dominates at long distance.  In the limit where 
 $T^3$ is shrunk to zero size,  the resulting Hermitian quantum matrix model is not well defined; the one-loop potential is unbounded from below for large eigenvalue separations.  For finite   $T^3 \times \R$, the eigenvalues can no longer run off to infinity; instead they spread over the dual 
 $\widetilde T^3$ uniformly.   
 We expect the center-symmetric phase to continue upon compactification, down to  
$T^3 \times S^1_\beta$ so long as  $\lambda_{1d}^{1/3} \beta \gg 1$.    In this domain,   
the center symmetry is intact, and in the large-$N$ limit, 
volume independence must  hold. 

In the high temperature limit,  $\lambda_{1d}^{1/3}\beta \ll 1$, 
  fermions decouple due to thermal mass,   eigenvalues clump, and 
 a  metastable bound state of eigenvalues exists. Let us assume 
 $|\beta\Delta x|, |\beta\Delta\alpha|\ll 1$, where $\Delta x = \max_{a,b}\{|\vec{x}_a-\vec{x}_b|\}$ 
and $\Delta \alpha = \max_{a,b}\{|\alpha_a-\alpha_b|\}$. 
Then the second term in the right hand side of \eqref{1loop_MQM} is negligible, due to the large thermal mass of fermions,   and the effective action becomes that of the zero-dimensional 
bosonic matrix model, 
\begin{eqnarray}
S_{\rm 1d,1-loop}^{-} \sim S_{\rm 0d,bos}[ x^{\mu}_{a}] 
&=&
2\sum_{ a< b}
\log |\vec x_{a}- \vec x_{b} |^2,
\end{eqnarray}
where we used the identification $x^4_a= \alpha_a$. 
This potential produces attraction between eigenvalues.  This model is studied extensively, and taking into account nonperturbative effects, 
a bound state  of eigenvalues  exists and it satisfies conditions  $|\beta\Delta x|, |\beta\Delta\alpha|\ll 1$.   In this domain, $(\mathbb Z_N)^4$ center symmetry is completely broken. 

Generalization to nonzero $m$ is straightforward and follows from the discussion of Sec.\ref{mainres} on $T^4$.   At sufficiently small $O(N^0)$  volume 
 such  that $m\ll \lambda_{1d}^{1/3}$,  the effect of mass is small,  and hence  
attraction in one-loop action is overwhelmed by nonperturbative  quantum fluctuations. In this domain, the $(\Z_N)^3$ center symmetry is intact.

\section{Lattice model and Monte-Carlo simulation} \label{lat}
\hspace{0.51cm}
In Sec.\ref{sec:0d pbc}, we explained on continuum $T^4$ and $T^3 \times S^1$  why the Eguchi-Kawai reduction holds 
for the QCD(Adj) at zero and finite temperature, based on perturbative-loop analysis, supplemented crucially with the estimates of nonperturbative  quantum fluctuations. 
Below, we study   the unitary matrix model and one-dimensional lattice model by using  Monte Carlo simulation. 

It is  hard to implement the thermal model  on a computer since  the temporal direction is not reduced.   Moreover,   in order to describe phenomena typical to the finite-temperature system, we need to take the  {\it effective spatial 
volume}\footnote{
The space size in its ordinary sense is fixed. 
By developing perturbation theory around a center symmetric background,  
say for simplicity with only one-dimension compact, we observe that both the lowest states and 
the spacing between the states is suppressed by a factor of $N$ and is given by 
$2 \pi/(LN)$ as opposed to the usual (center-broken)  KK-spectrum  where the level spacing is $2 \pi/L$ . 
In other words, to all orders in perturbation theory, 
the effective space size is enhanced into $L_{\rm eff} = LN$. 
(See the discussion in \cite{PU09} for QCD(Adj) and related discussions in a review 
\cite{Das:1984nb} for the quenched Eguchi-Kawai model (QEK) and the twisted Eguchi-Kawai model (TEK).)  
This is the perturbative essence of volume independence, 
and in the sense of neutral sector observables, 
decompactification limit can be reached at $N \rightarrow \infty$ while keeping $L$ fixed. \label{fn:ess}
} sufficiently large compared to $N_t$.
As for a lattice model with an isotropic lattice spacing, 
roughly speaking, we need to take the spatial lattice size $N_s$
to be twice larger than the temporal one $N_{t}$, $N_s\gtrsim 2N_t$. 
Since the  $N_s^{\rm eff}$  is related to the matrix size $N$ for 
the large-$N$ reduced models in the sense described in footnote \ref{fn:ess}, 
we need to take $N$ large enough to satisfy this condition.  When we use
the original Eguchi-Kawai reduction for  the spatial directions, 
finite-$N$ correction behaves as $1/N$ rather than $1/N^2$, as we will show numerically.  
This fact indicates that  the effective lattice size scales as $N_s^{\rm eff} \sim N^{1/3}$,  
and it is not practical for  numerical simulations.
Therefore, to gain  more effective spatial  volume, 
we impose  a twisted boundary condition on the spatial directions. For the latter, 
 the effective volume  is more enhanced,   and volume independent domain can be reached more quickly.  We call the QCD(Adj) with the 
twisted boundary condition as TQCD(Adj). 
 The introduction of TQCD(Adj), which is algorithmically more convenient, is our main improvement over the one-site model of 
 Ref.\cite{BS09}. 

To explain the efficiency of the TQCD(Adj), we start with the zero temperature case, that is, the 
single-site model with periodic boundary condition along the temporal direction on fermions. 
Then we apply the twist to the finite-temperature case, that is, the 
one-dimensional lattice model with an antiperiodic boundary condition on fermions along the temporal 
direction, and then see the Eguchi-Kawai reduction holds in this case. 
Especially we show this reduced model at finite-temperature describes the confinement/deconfinement transition.

\subsection{Single-site theories}
\subsubsection{QCD(Adj) on $1^4$ lattice }\label{sec:single site_nontwist}
\hspace{0.51cm}
The single-site matrix model  can be  obtained from a four-dimensional lattice gauge theory  
 by reducing the number of lattice sites to one  in all directions \cite{BS09}.  
  The action is 
\begin{eqnarray}
S_{0d}
=
-2b N
Re\ Tr \left(
\sum_{\mu<\nu}
V_\mu V_\nu V_\mu^\dagger V_\nu^\dagger
\right)
+
S_F,  
\end{eqnarray}
where $V_\mu\ (\mu=1,2,3,4)$ corresponds to the link variable in the four-dimensional theory. 
The inverse 't Hooft coupling constant $b$ should be chosen appropriately 
depending on the lattice spacing $a$.  
The fermionic part $S_F$ is obtained as  dimensional reduction 
of the Wilson-Dirac fermion term 
\begin{eqnarray}
S_F
=
\displaystyle{\sum_{f=1}^{N_f^D}} \left(
\bar{\psi}_f\psi_f
-
\kappa\sum_{i=1}^3\left\{
\bar{\psi}_f(1-\gamma_\mu)V_\mu\psi_f V_\mu^\dagger
+
\bar{\psi}_f(1+\gamma_\mu)V^\dagger_\mu\psi_f V_\mu 
\right\} \right). 
\end{eqnarray} 
The hopping parameter $\kappa$ can be expressed as 
\begin{eqnarray}
\kappa=\frac{1}{8+2am_0}, 
\end{eqnarray}
where $m_0$ is the bare mass.

This action has a $({\mathbb Z}_{N})^4$ center symmetry 
\begin{eqnarray}
V_\mu
\to
e^{2\pi in_{\mu}/N}V_\mu\qquad (n_{\mu} =0,1,\cdots,N-1). 
\end{eqnarray}    
If this symmetry is not broken,  then the   model is equivalent to the translationally invariant subsector of  lattice theory  with an arbitrary number of sites, including an infinite lattice limit.

Although detailed analytic evaluation of  one-loop effective potential  depends  on  the choice of  lattice fermions,\footnote{Here, we use Dirac-Wilson fermion with Wilson parameter $r=1$, 
\cite{HN09} uses overlap fermions, and also discusses naive fermions.}
 intuitively,  the absence of the center-symmetry breaking phase 
follows closely the discussion on continuum $T^4$. The discussion on $T^4$ can easily be generalized to the $1^4$ lattice. The role of the compactification scale  $L$  is replaced by the lattice spacing $a$ in the one-site model.  The main lesson that we learn is that  the center symmetry on the $1^4$ lattice model is in fact much more robust than the center on continuum $T^4$.

The one-loop action  for the one-site theory in the classical background of commuting Wilson lines  $[V_{\mu}, V_{\nu}] =0$ where   $V_\mu={\rm diag}(e^{i\theta_\mu^1},\cdots,e^{i\theta_\mu^N})$ is given by 
\begin{eqnarray}
S_{\rm 1-loop}
=
2\sum_{a<b}\log\Big[
\frac{4}{a^2}\sum_{\mu=1}^4\sin^2\Big (\frac{\theta_\mu^{ab}}{2}\Big)
\Big]
-
4N_f^D\sum_{a<b}\log \left( 
\frac{1}{a^2}\sum_{\mu=1}^4\sin^2\theta_\mu^{ab}
+ \Big( m_0+ \frac{2}{a}\sum_{\mu=1}^4\sin^2\Big(\frac{\theta_\mu^{ab}}{2}\Big)
\Big)^2
\right), 
\nonumber\\
\label{1loop-l}
\end{eqnarray}
The first term is  induced by gauge fluctuations and leads to eigenvalue attraction \cite{BHN82}. 
Geometrically,  
\begin{equation}
  {\cal P}_{\mu}^{ab} \equiv   
\frac{2}{a} \left|\sin \left( \frac{ \theta_{\mu}^{a} - \theta_\mu^{b}}{2}  \right)  \right| = \frac{2}{a} |e^{i   \theta_{\mu}^{a}} - e^{i   \theta_{\mu}^{b}}  | 
\end{equation}
  is the separation 
 between two  eigenvalues of the Wilson line in the $\mu$ direction.  $  {\cal P}^2 \equiv \sum_{\mu=1}^{4} ({\cal P}_{\mu}^{ab})^2$ is  the  spectrum of massive gauge fluctuations ($W$ bosons),  familiar from the usual 
D-brane pictures as the spectrum of open strings ending on branes, where eigenvalues   of Wilson line are identified with branes.
The second term, proportional to $N_f^D$, is induced by fermionic fluctuations, and 
is equal to $ -4N_f^D   \sum_{a<b}    \; \log \left[  M_f^2[\theta_{\mu}^{ab}] \right]  $
where  $M_f^2[\theta_{\mu}^{ab}]$ is the spectrum of fermions.

Eq.(\ref{1loop-l})  may be rewritten in a form mimicking the continuum expression on $T^4$ given in  (\ref{smallT4}): 
\begin{equation}
S_{ \rm 1-loop}  [\theta_{\mu}^{ab}] =  2  \sum_{a<b}    \;  \log \left[ \sum_{\mu=1}^4  ({\cal P}^{ab}_{\mu})^2   \right]  -4N_f^D  \sum_{a<b}    \; \log \left[  
\sum_{\mu=1}^4    ({\cal P}^{ab}_{\mu})^2   +     \sum_{\mu < \nu}   \frac{a^2  ({\cal P}^{ab}_{\mu})^2  ({\cal P}^{ab}_{\nu})^2   }{2 (1+ m_0a )} 
 +    \frac{m_0^2}{ (1+ m_0a)}  
 \right].
\label{1-site}
\end{equation}
where we have subtracted a holonomy-independent constant term.  
This expression can be Poisson resummed  and be written in terms of  Wilson lines as 
\begin{equation}
S_{\rm 1-loop} [\theta_{\mu}^{ab}] = \sum_{a<b} \sum_{ \vec n \in \Z^4\setminus\{{\bf 0}\} }
  e^{i \vec \theta^{ab} \cdot \vec n} P_{\vec n}(m_0a) \; , 
\end{equation} 
as discussed in the Appendix~\ref{Poisson}. 
The only difference with respect to the continuum expression on $T^4$ is that 
\begin{equation}
P_{\vec n}^{\rm 1site}(0) = e_{\vec n}   P_{\vec n}^{T^4}(0)\,,
\end{equation}
   where 
 $e_{\vec n} $  is an enhancement factor of a one-site model over continuum $T^4$, for both bosonic and fermionic contribution. In general, due to peculiarities of the   dispersion relation of the one-site model,  the center stability is further enhanced on the one-site model with respect to continuum 
 $T^4$. Asymptotically, for $|{\vec n}|\gg 1$,   $e_{\vec n} \rightarrow 1 $ as expected on physical grounds, by just inspecting the dispersion relations. 
 
 The fermionic contribution to $e_{\vec n} $ is numerically sizeable and has interesting implications. 
   For example, on $T^4$, start with $mL=\infty $. Following the discussion of the Appendix~\ref{Poisson},  the singly-winding  Wilson lines do get stabilized (in perturbation theory) at $mL=2.027$, whereas for the one-site model, 
the same phenomena take place at   $m_0 a=9.3$.  This is due to the fact that in the domain of  
 heavy fermion bare mass,   
the last term in (\ref{1-site}), which may roughly be viewed as an  ``effective mass" $m_{\rm eff}^2$, is suppressed with respect to the bare mass, $m_{\rm eff} \sim \sqrt{m_0 /a}$.  
For $ N_f^D =1$ theory,   the fermions' contribution dominates, leading to eigenvalue repulsion, and unbroken center symmetry for the theory defined on a $1^4$ lattice, or EK reduction of QCD(Adj) to a single-site lattice.

 If one-loop perturbation theory was the whole description,  this would be the transition to a $(\Z_2)^4$ restored phase. However, as explained above, many phases can coexist and it is beyond perturbation to determine 
which phase is chosen in the end. Interestingly, the point where the center symmetry (partially) restores 
in simulation is very close to the prediction by perturbation theory $\kappa\approx 0.037$.   
This is approximately the value of $\kappa$ where Ref.\cite{BS09} observed the full center restoration, 
not just $\Z_2$. In Fig.~\ref{fig:Hysteresis_Nf1S1N25B050}, we can also see the restoration around the same 
value. Although we could not see partial breaking clearly by measuring the observable employed in Ref.\cite{BS09}, 
the discrepancy between $\langle |W|\rangle$ in large-$\kappa$ start and the one in small-$\kappa$ start 
may indicate the existence of a partial breaking phase.

\subsubsection{TQCD(Adj) at zero temperature}
\hspace{0.51cm}

In compactified QCD(Adj), as we will see explicitly from simulations, the 
finite-$N$  corrections turn out to be order $1/N$,   as opposed to the perturbative  expectation  on $\R^4$ \cite{'tHooft:1973jz}. As explained in the results 
section of  Sec.\ref{res}, there are  two plausibly  related  explanations for this behavior. One is related to the discussion of effective volume in the reduced model.  In the reduced model, $N$ serves the role of an emergent spacetime volume, at least in a perturbative description in finite volume  around  a center-symmetric configuration. 
Finite-$N$ corrections should  scale as finite-volume corrections. However, what is not always clear is the factor $N^p$  via which volume enhancement takes place $V_{\rm eff} \sim N^pV$,~\footnote{Of course, for QEK-like configuration, $p=1$ and for TEK-like configurations, 
$p=2$, see, for example, \cite{Das:1984nb}. However, these deductions are in perturbation theory around particular backgrounds, and the  determination of $p$ is likely non-perturbative.} 
and   $p$   may in fact be determined nonperturbatively. 

In compact space,  one cannot gauge away zero-momentum modes, and these modes are crucial in studying perturbation theory in finite volume.  In perturbation theory, the spectrum of the theory relies on the background for the Wilson lines. If theory has massless adjoint fermions, there will also be fermionic zero modes in the spectrum. 
 Typically, there are order $N$  light or massless bosonic   and  order $N$ fermionic zero modes, which may generate nonperturbative $1/N$ effects \cite{BS10}.

Both problems  can simultaneously be solved   and  $1/N$ corrections can systematically be improved  by using boundary conditions which cannot be obeyed by either bosonic or fermionic (if there are any) zero modes.   This can be done by using  the twisted boundary conditions of  't Hooft  \cite{'tHooft:1979uj}. 
This idea is, of course, not new, and is used  by Witten in  Ref.\cite{Witten:1982df} to lift the zero modes in ${\cal N}=1$ 
super Yang-Mills theory in the context of supersymmetric theories on $T^3 \times \R$, and   by   Gonzalez-Arroyo and Okawa \cite{GAO82}  in the context of large-$N$ 
reduced models.

Our main observation can be summarized by using the following pedagogical exercise. (The generalization to the theories that we use in simulations is straightforward. The prescription given below  works equally well on lattice and continuum.) 
 Let $\Phi (x_1, x_2, x_3, x_4) $ denote either a unitary gauge field   or an adjoint fermion  field.  We impose  the following generalized    boundary conditions on fields
\begin{eqnarray}
&&\Phi (\ldots, x_{\mu}+ L, \ldots  )= B_{\mu} \Phi ( \ldots, x_{\mu} , \ldots  ) B_{\mu}^{\dagger}\,\, .
\end{eqnarray}
 For the particular case of one-site matrix models,  we can set $\Phi(x_1, x_2, x_3, x_4)  
 = \Phi =$constant. For the theory on the $1^4$ lattice, 
 we consider two  choices for $B_{\mu}$. 
\begin{eqnarray}
&&{\rm pbc}: \qquad  \;\;\;\;\;  B_{\mu}=1_N \cr 
&&{\rm Twist}:  \qquad B_{1}=C_{\sqrt N}  \otimes 1_{\sqrt N} \; , \;  B_2= S_{\sqrt N}  \otimes 1_{\sqrt N} \; ,\;  B_{3}=1_{\sqrt N} \otimes  C_{\sqrt N} \; , \;   B_4=1_{\sqrt N} \otimes 
 S_{\sqrt N}   \; \qquad 
\label{bcs}
\end{eqnarray}
where $C_{\sqrt{N}}$ and $S_{\sqrt{N}}$ are $\sqrt{N}\times\sqrt{N}$ (noncommuting) 
clock and shift matrices obeying $C_{\sqrt{N}}S_{\sqrt{N}} =  e^{-i \frac{2\pi}{{\sqrt{N}}}} S_{\sqrt{N}} C_{\sqrt{N}}$. A particular representation is 
\begin{eqnarray}
C_{\sqrt{N}}
=
diag(1,\omega,\omega^2,\cdots,\omega^{\sqrt{N}-1}), 
\qquad
S_{\sqrt{N}}
=
\left(
\begin{array}{ccccc}
0 & 1 & 0 & \cdots & 0\\
0 & 0 & 1 & \cdots & 0\\
\vdots & \vdots & \vdots & \ddots & \vdots\\
0 & 0 & 0 & \cdots & 1\\
1 & 0 & 0 & \cdots & 0
\end{array}
\right),  
\end{eqnarray}
where $\omega=\exp(2\pi i/\sqrt{N})$.

The first case in (\ref{bcs}) is the original   QCD(Adj) with  periodic boundary conditions 
and does not lift any zero or light modes associated with holonomy, or  fermions. In this case, 
finite-$N$ corrections turn out to be largest, of order $1/N$. 

The second case in (\ref{bcs}) is a  twist of    QCD(Adj).   
The twist lifts all possible zero or light modes from the spectrum.  In this case, 
finite-$N$ corrections turns out to be  of order $1/N^2$.

The action of the theory with twisted boundary conditions can be turned into a theory with periodic boundary conditions and an action with an insertion of 't Hooft flux. This is our definition of  ``twisted" QCD(Adj) [or TQCD(Adj)]: 
\begin{eqnarray}
S_{0d}
=
-2b N
Re\ Tr \left(
\sum_{\mu<\nu}
Z_{\mu\nu}
V_\mu V_\nu V_\mu^\dagger V_\nu^\dagger
\right)
+
S_F,    
\end{eqnarray}
where $Z_{\mu\nu}$ is the twist factor. Geometrically,   $Z_{\mu\nu}$ is associated with the 't Hooft flux passing 
through  the $(\mu\nu)$ plaquette.  
Here we adopt ``symmetric twist"
\begin{eqnarray}
Z_{\mu\nu}
=
Z_{\nu\mu}^\ast
=
e^{2\pi i/\sqrt{N}} 
\qquad(\mu<\nu) \; .
\end{eqnarray}
 As the area enclosed   by fermionic ``plaquette" terms is zero,  the flux passing through it   is zero. Thus, fermionic action is   unaltered.   This procedure, apart from helping   QCD(Adj) algorithmically, also cures the global instability 
\cite{BNSV06,AHHI07,TV06} of the TEK model. 
   We numerically 
compare the behaviors of  finite-$N$  corrections for QCD(Adj) and 
TQCD(Adj) below.

\subsubsection{Numerical results for $N_f^D=1$}
\hspace{0.51cm}
We now discuss  the Monte Carlo results for QCD(Adj) and TQCD(Adj) at zero temperature. 
We restrict our analysis to the case with a single Dirac fermion in adjoint representation.\footnote{
We implemented the rational hybrid Monte Carlo (RHMC) algorithm \cite{KHS98} with the multi-mass conjugate gradient (CG) solver \cite{Jegerlehner:1996pm}. 
Numerical coefficients in the rational approximation 
necessary for the RHMC simulation was obtained by using the simulation code provided at \cite{Remez}.}
 
\begin{figure}[htbp]
\begin{tabular}{cccc}
\begin{minipage}{0.50\hsize}
\begin{center}
\scalebox{0.40}{
\rotatebox{-90}{
\includegraphics{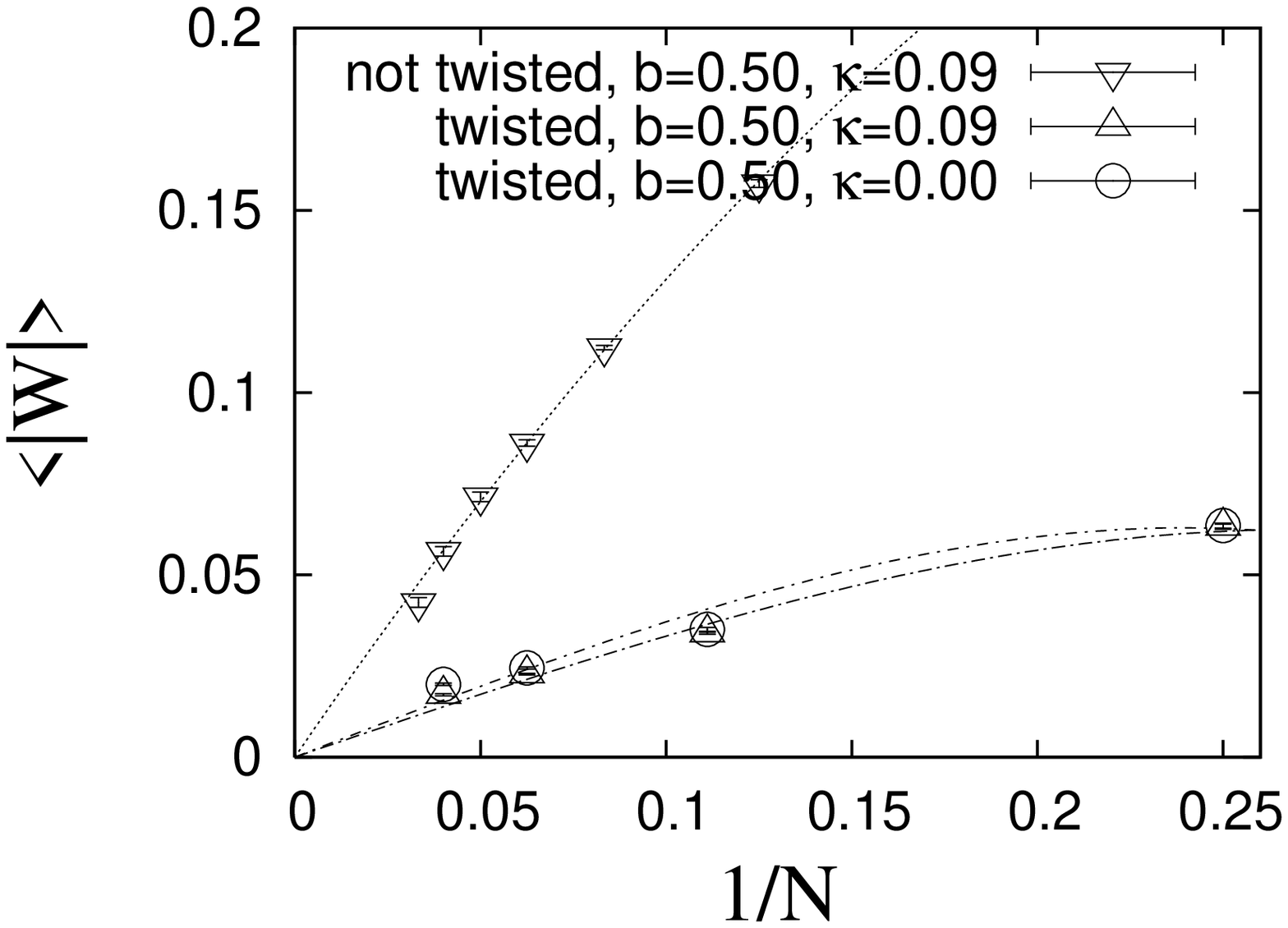}}}
\caption{ 
Expectation values of the Wilson loop in QCD(Adj) and TQCD(Adj) at $b=0.50, \kappa=0.09$ and $\kappa=0$ 
(bosonic twisted Eguchi-Kawai model). 
Fitting curves are of the form $c/N+d/N^2$ for the former and $c'/N+d'/N^3$ for the latter. 
TQCD(Adj) at $b=0.50, \kappa=0.09$ and $\kappa=0$ agree quite well, 
as expected because $\kappa=0.09$, corresponds to a quite heavy fermion.    
}\label{fig:B050K009WilsonLoop}
\end{center}
\end{minipage}
\begin{minipage}{0.4\hsize}
\end{minipage}
\begin{minipage}{0.4\hsize}
\end{minipage}
\begin{minipage}{0.45\hsize}
\begin{center}
\vspace{-1.5cm}
\scalebox{0.40}{
\rotatebox{-90}{
\includegraphics{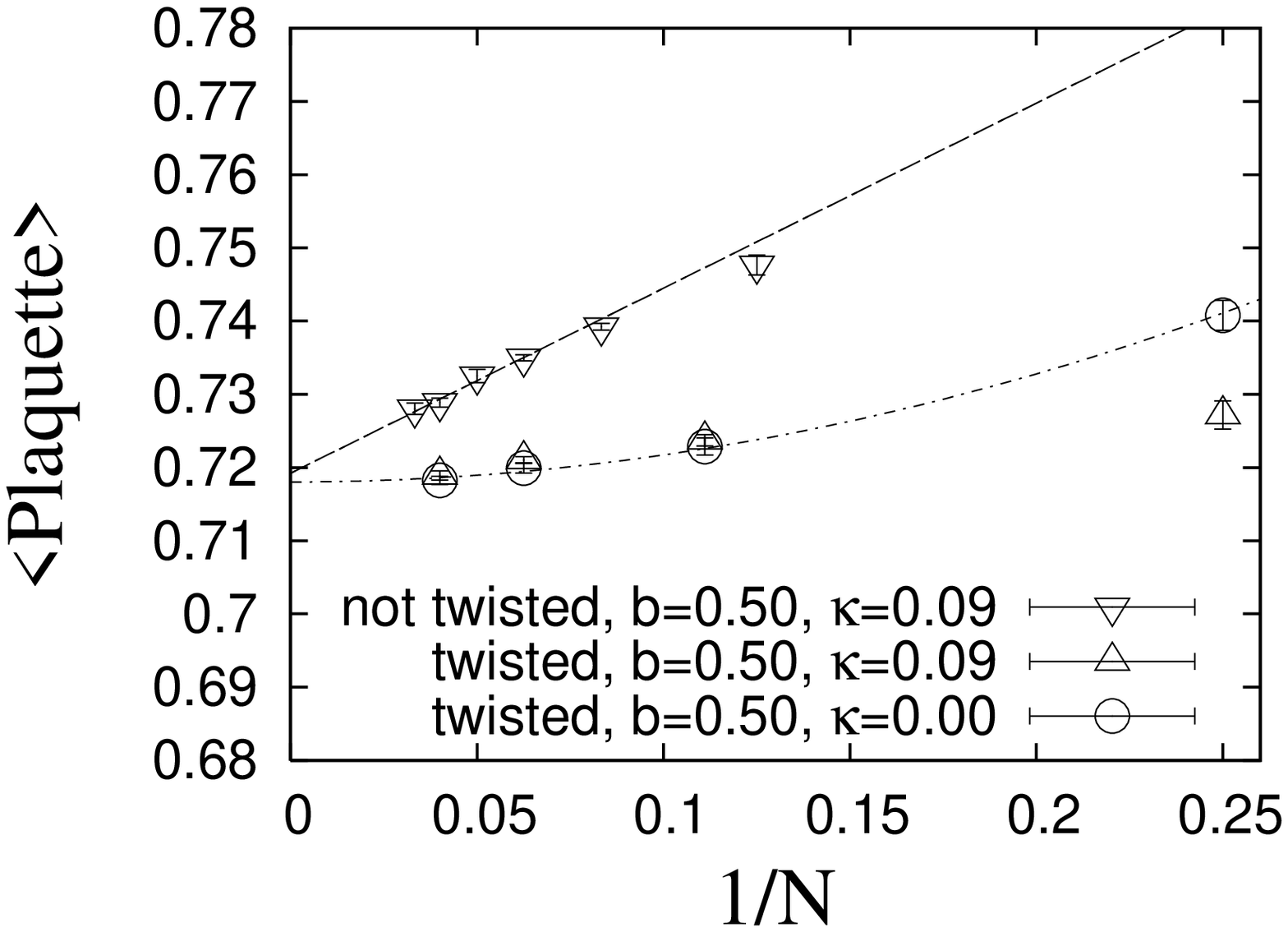}}}
\caption{ 
Expectation values of the plaquette in QCD(Adj) at $b=0.50, \kappa=0.09$ 
and TQCD(Adj) at $b=0.50, \kappa=0.09$, $\kappa=0$. 
The $1/N$ correction is of order $1/N$ for the QCD(Adj) and $1/N^2$ for TQCD(Adj).   
}\label{fig:B050K009}
\end{center}
\end{minipage}
\end{tabular}
\end{figure}

In Fig.~\ref{fig:B050K009WilsonLoop},  the  expectation value for the absolute value of the Wilson loop  (averaged over 
all directions) 
\begin{eqnarray}
|W|
\equiv
\frac{1}{4}\sum_{\mu=1}^4|V_\mu|,
\end{eqnarray} 
in the QCD(Adj) and the TQCD(Adj) at zero temperature
is plotted.\footnote{We use absolute value of the Wilson line  operator in small volume to distinguish  a center-symmetric  saddle point from a  multi-saddle configurations  
for  which $\langle W \rangle$  is non-vanishing at each saddle, but vanishes due to phase averaging  over all saddles (which is permitted in quantum theory due to tunneling). Multi-saddle 
configurations,  in the large-$N$ limit, lead to spontaneously breaking of the    center-symmetry, whereas a 
center symmetric saddle continues to respect the center symmetry. }
For both the QCD(Adj) and TQCD(Adj), $\langle |W|\rangle$ is of order $1/N$ and hence the $(\Z_N)^4$ symmetry is unbroken. 
(As already shown in \cite{BS09}, 
it is unbroken in a rather large parameter region.)  
The extent of the next-to-leading correction 
is not clear from this plot; we fit it by  
$\langle |W|\rangle\sim c/N+d/N^2$ for QCD(Adj) and 
$\langle |W|\rangle\sim c'/N+d'/N^3$ for TQCD(Adj), 
where $c,d,c',d'$ are constants.

In Fig.~\ref{fig:B050K009}, expectation values of the plaquettes are  plotted.  
From this plot, the finite-$N$ correction for the QCD(Adj) turns out to be of order $1/N$. 
On the other hand, the finite-$N$ correction for the TQCD(Adj) is of order $1/N^2$ as expected. 

\begin{figure}[htbp]
\begin{center}
\scalebox{0.25}{
\rotatebox{-90}{
\includegraphics{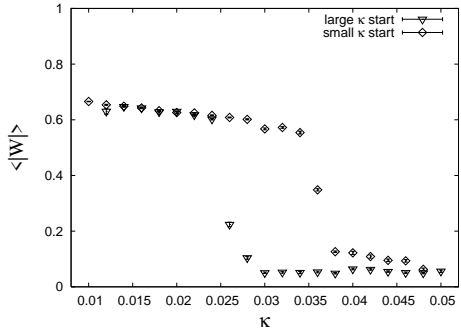}}}
\caption{ 
Expectation values of the Wilson loop $\langle|W|\rangle$ in QCD(Adj) at $b=0.50$, $N=25$. 
Clear hysteresis can be seen.   
}\label{fig:Hysteresis_Nf1S1N25B050}
\end{center}  
\end{figure}

In Fig.~\ref{fig:Hysteresis_Nf1S1N25B050}, expectation values of Wilson lines  $\langle|W|\rangle$ 
near the phase transition are shown. 
The argument in Sec.\ref{sec:single site_nontwist} suggests the transition is of first order 
because several phases coexist. To confirm, we studied and observed hysteresis. 
We started simulation at $\kappa=0.01$ (small-$\kappa$ start) and 
$\kappa=0.05$ (large-$\kappa$ start), and gradually increased/decreased the value of $\kappa$. 
At each point, we collected 500 -- 2000 samples, which is enough to evaluate 
the expectation values. As can be seen from the plot, there is a clear hysteresis. 
Thus we conclude the transition is indeed of first order.
 
We also studied the distribution of $Tr(V_\mu V_\nu)/N$ as in \cite{BS09}, 
and did not find partial breaking of the center symmetry. However, 
around $\kappa=0.04$ in Fig.4, there is some discrepancy between 
$\langle|W|\rangle$ in the large-$\kappa$ start and that in the small-$\kappa$ start. 
Hysteresis continues with a smaller gap within $0.0375 < \kappa<0.0475$.
This may be interpreted as a partial breakdown of center symmetry
\cite{SSpc} , where the unbroken phase coexists with a partially
broken phase.

\subsection{One-dimensional lattice: Reduced model at finite temperature}
\hspace{0.51cm}
We first introduce 
a  one-dimensional lattice formulation  corresponding to the large-$N$ reduced model at 
finite temperature and then apply two different types of twists to this model. 
 An implication of volume independence is that one can study 
 confinement/deconfinement  transition on $\R^3 \times S^1_\beta$ on the equivalent unitary matrix model, corresponding to a $1^3 \times N_t$ lattice so long as $(\Z_N)^3$ center symmetry associated with the spatial cycles  is not spontaneously broken. We indeed observe  
 the confinement/deconfinement 
transition  in the reduced model.  For the comparision of $1/N$ corrections, 
we  also  plot some numerical results for the QCD(Adj) (without twist) at finite temperature.   
\subsubsection{(T)QCD(Adj) at finite temperature}
On the $1^3 \times N_t$ lattice or $T^3 \times S^1_\beta$ continuum  formulations, the relation between twisted boundary conditions, zero (or light) modes,   and suppression of finite-$N$ effects can systematically be  studied.

 Let $\Phi (t, x,y,z) $ denote either a unitary gauge field   or an adjoint fermion  field. 
We impose,  the following generalized    boundary conditions on fields
\begin{eqnarray}
&&\Phi (t, x+ L , y, z)= A \Phi (t,x, y, z) A^{\dagger}\, , \cr
&&\Phi (t,x, y+L, z)= B \Phi (t,x, y, z) B^{\dagger}\, , \cr
&&\Phi (t, x, y, z+ L)=  \Phi (t, x, y, z)\,.  
\end{eqnarray}
For the particular case of matrix models, 
we can set $\Phi (t,x, y, z) = \Phi(t)$. 
We consider three choices for 
$A$ and $B$. 
\begin{eqnarray}
&&{\rm pbc}: \qquad  \;\;\;\;\;  A=1_N \; , \qquad B= 1_N \cr 
&&{\rm Twist \;1}:  \qquad A=C_{\sqrt N}  \otimes 1_{\sqrt N} \; ,  \qquad B= S_{\sqrt N}  \otimes 1_{\sqrt N} \cr
&& {\rm Twist \; 2}: \qquad A=C_N \; ,  \qquad B= S_N 
\label{bcs2}
\end{eqnarray}
where $C_N$ and $S_N$ are $N \times N$ 
clock and shift matrices 
defined earlier. 

The first case in (\ref{bcs2}) is the original   QCD(Adj) with  periodic boundary conditions 
and does not lift any zero or light modes associated with holonomy, or  fermions. In this case, 
finite-$N$ corrections turn out to be  of order $1/N$. 

The second case in (\ref{bcs2}) is a partial twist of    QCD(Adj). The twist lifts a $1/\sqrt N$ fraction of bosonic light modes and fermionic zero modes (in cases where fermions are light). This can  be seen by explicitly solving the boundary conditions.  In this case, 
finite-$N$ corrections turn out to be  of order $1/N^{3/2}$. This is explained below after 
introducing the twisted model.

The third case in (\ref{bcs2}) is a  twist of    QCD(Adj), which  lifts all  bosonic light modes and fermionic zero modes.   This can  be seen by explicitly solving the boundary conditions. In this case, finite-$N$ corrections seem to be rather tame.

We study the theory reduced to a one-dimensional SU(N) unitary matrix model 
\begin{eqnarray}
S_{lat}
=
-2b N\sum_t 
Re\ Tr\left(
\sum_i
U_t(t)V_i(t+1)U_t^\dagger(t)V_i^\dagger(t)
+
\sum_{i<j}Z_{ij}
V_i(t)V_j(t)V_i^\dagger(t)V_j^\dagger(t)
\right)
+
S_F,  
\end{eqnarray}
where $V_i(t)\ (i=1,2,3)$ corresponds to the link variable along the spatial direction 
in the four-dimensional theory.
For the usual nontwisted model $Z_{ij}=1\,$ and $Z_{ii}=0$, 
while, for twisted models, $Z_{ij}$ is given by\footnote{In simulations, we use symmetric twist  which is more efficient. The above boundary conditions can be modified to incorporate the symmetric twists.}
\begin{eqnarray}
&&{\rm Twist \; 1}:  Z_{ij} = Z_{ji}^\ast
=
e^{2\pi i/\sqrt{N}}
\qquad(i<j), \cr 
&&{\rm Twist \; 2}:  Z_{ij} = Z_{ji}^\ast
=
e^{2\pi i/N}
\qquad(i<j). 
\end{eqnarray}
The number of sites $N_t$ is related to the temperature $T$
by $\beta=1/T=aN_t$.  
The fermionic part $S_F$ is given by 
\begin{eqnarray}
S_F=\sum_{f=1}^{N_f^D}
\bar{\psi}_f D_W^{(f)}\psi_f, 
\end{eqnarray}
where $D_W$ is the usual Wilson-Dirac operator. 
The explicit form of $S_F$ after reducing the spatial directions is 
\begin{eqnarray}
S_F
&=&
\sum_t\biggl(
\bar{\psi}(t)\psi(t)
-
\kappa\sum_{i=1}^3\left\{
\bar{\psi}(t)(1-\gamma_i)V_i(t)\psi(t)V_i^\dagger(t)
+
\bar{\psi}(t)(1+\gamma_i)V^\dagger_i(t)\psi(t)V_i(t)
\right\}
\nonumber\\
& &
\qquad
-
\kappa\left\{
\bar{\psi}(t)(1-\gamma_t)U_t(t)\psi(t+1)U_t^\dagger(t)
+
\bar{\psi}(t)(1+\gamma_t)U^\dagger_t(t-1)\psi(t-1)U_t(t-1)
\right\}
\biggl). 
\nonumber\\
\end{eqnarray}  

We impose  antiperiodic boundary condition 
for the fermions on the $S^1_\beta$ and generalized  boundary conditions  given in (\ref{bcs2}) 
on reduced directions.  
The action has a global center symmetry, which we split for convenience as 
$(\Z_N)^3 \times  \Z_N$, 
\begin{eqnarray}
V_i(t)
\to
e^{2\pi i/N}V_i(t), \qquad  (U_{t=1}\ldots U_{t=N_t}) \rightarrow e^{2\pi i/N} 
 (U_{t=1}\ldots U_{t=N_t}).  
\end{eqnarray}    
If the $(\Z_N)^3$ symmetry is not spontaneously broken, 
then this model is equivalent to the one in the infinite spatial volume lattice. 
This implies the phase transition of infinite volume theory, associated 
with the realization of the temporal $\Z_N$ factor, 
can be studied by using the  unitary matrix model.

The relation between the twists, number of light modes, and the observed form of the 
finite-$N$ corrections is
\begin{equation}
\begin{array}{|c|c|c|}
\hline
{\rm Twist} & {\rm   Zero\, modes} & {\rm finite-} N \;  {\rm corr.} \cr 
\hline
{\rm none} & N & 1/N \cr
\hline
e^{i\frac{ 2 \pi}{\sqrt N} } & \sqrt N &  1/N^{3/2}  \cr 
\hline
e^{i \frac{ 2 \pi}{N} } & {\rm none} &  1/N^{2}  \cr
\hline
\end{array} 
\end{equation}

There is a nice geometric interpretation for twist 1 in terms of a classical  background, and the foliation of  noncommutative plane. This gives, in perturbation theory, that effective volume
should scale as  $V\sim N^{3/2}$. However, for twist 2, there is no classical background solution. Because of strong quantum fluctuations, a classical background cannot be written.   Of course, this is not a concern.

The emergence of $V\sim N^{3/2}$ in the case of twist 1 can be explained in perturbation theory 
as follows: 
For simplicity, let us  again consider the  twist 
$Z_{12}
=
e^{2\pi i/\sqrt{N}}
$, $Z_{13}=Z_{23}=1$. 
A natural candidate of the ground state is 
\begin{eqnarray}
V_1=C_{\sqrt{N}}\otimes \textbf{1}_{\sqrt{N}}, \quad
V_2=S_{\sqrt{N}}\otimes \textbf{1}_{\sqrt{N}}, \quad
V_3= \textbf{1}_{\sqrt{N}}\otimes C_{\sqrt{N}}. 
\end{eqnarray}

This is because this configuration satisfies $V_iV_j=Z^{\ast}_{ij}V_jV_i$, 
and at the same time eigenvalues spread as uniformly as possible. 
This configuration  keeps the $(\Z_{\sqrt{N}})^3$ subgroup of the center symmetry, which 
is enough for the Eguchi-Kawai reduction to work. 
Then, along $V_1$ and $V_2$ directions  $\sqrt{N}$ sites arise as a fuzzy torus, similar
to TEK, and $\sqrt{N}$ sites emerges along the $V_3$ direction similar  to QEK. 
So this configuration corresponds to  $V^{\rm eff} \sim N^{3/2}$ lattice sites. 
If one views finite-$N$ corrections around $N=\infty$ analogous of finite-volume corrections in compactified theories,  then, 
it is expected that finite-$N$ corrections in the 
one-dimensional reduced model should be a power series expansion in  $1/N^{3/2}$.

\subsubsection{Monte Carlo simulation for $N_f^D=1$}\label{sec:MCFT}
\hspace{0.51cm} 
Here we show the Monte Carlo results for the one-dimensional lattice model  for $N_f^D=1$.   
The numerical algorithm adopted is the same as the one in the previous subsection.

\begin{figure}[htbp]
\begin{tabular}{ccccc}
\begin{minipage}{0.33\hsize}
\begin{center}
\rotatebox{-90}{
\scalebox{0.33}{
\includegraphics{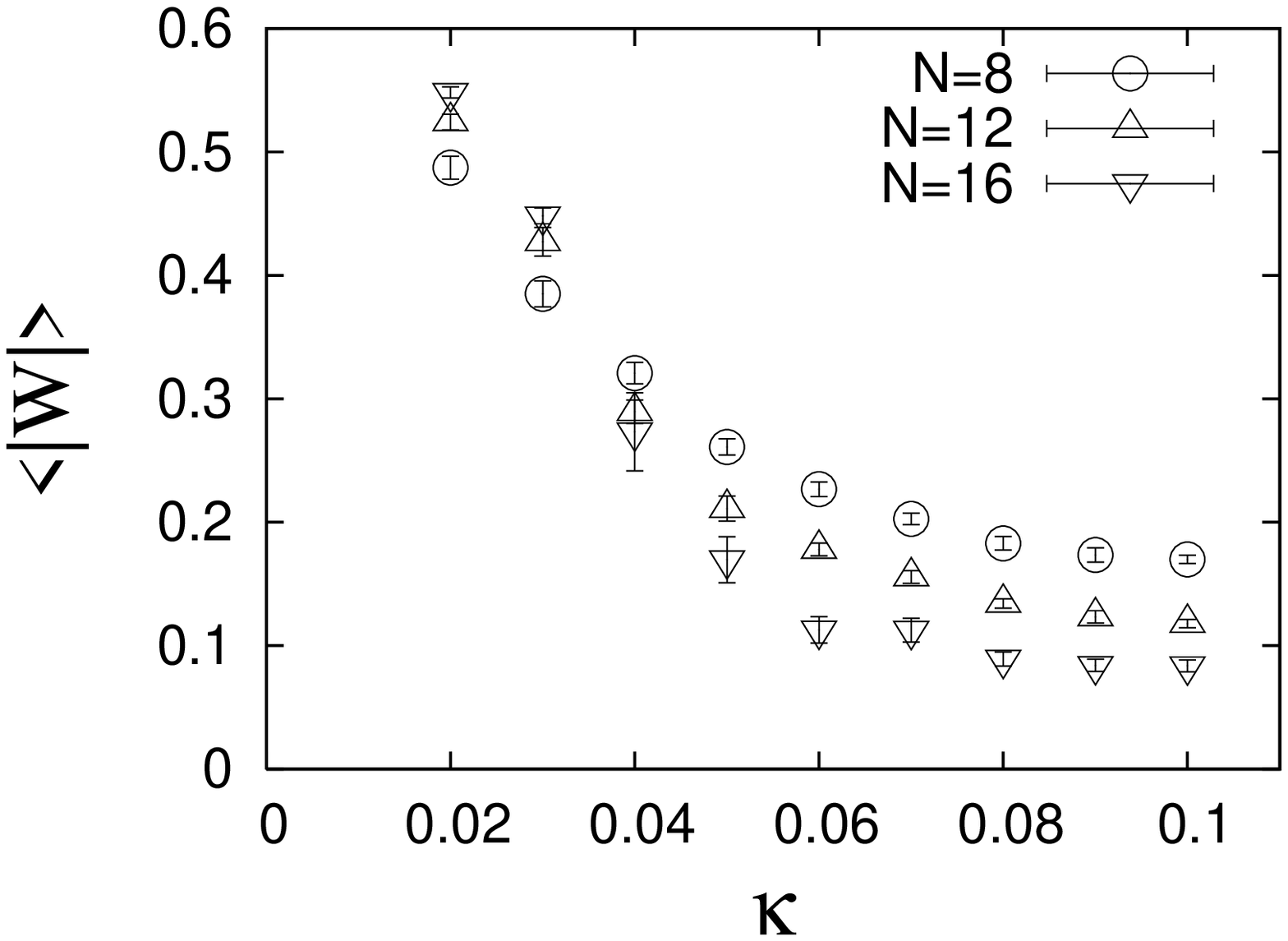}}}
\caption{
Expectation value of the spatial Wilson line  in QCD(Adj) (without twist)
at $b=0.50$. 
When the quark is lighter, the Wilson line  becomes zero at large-$N$; 
that is, $(\Z_N)^3$ center symmetry is not broken. 
}
\label{fig:Wilson_apbc_kappa_dependence}
\end{center}
\end{minipage} 
\begin{minipage}{0.33\hsize}
\end{minipage}
\begin{minipage}{0.33\hsize}
\begin{center}
\vspace{-2cm}
\rotatebox{-90}{
\scalebox{0.33}{
\includegraphics{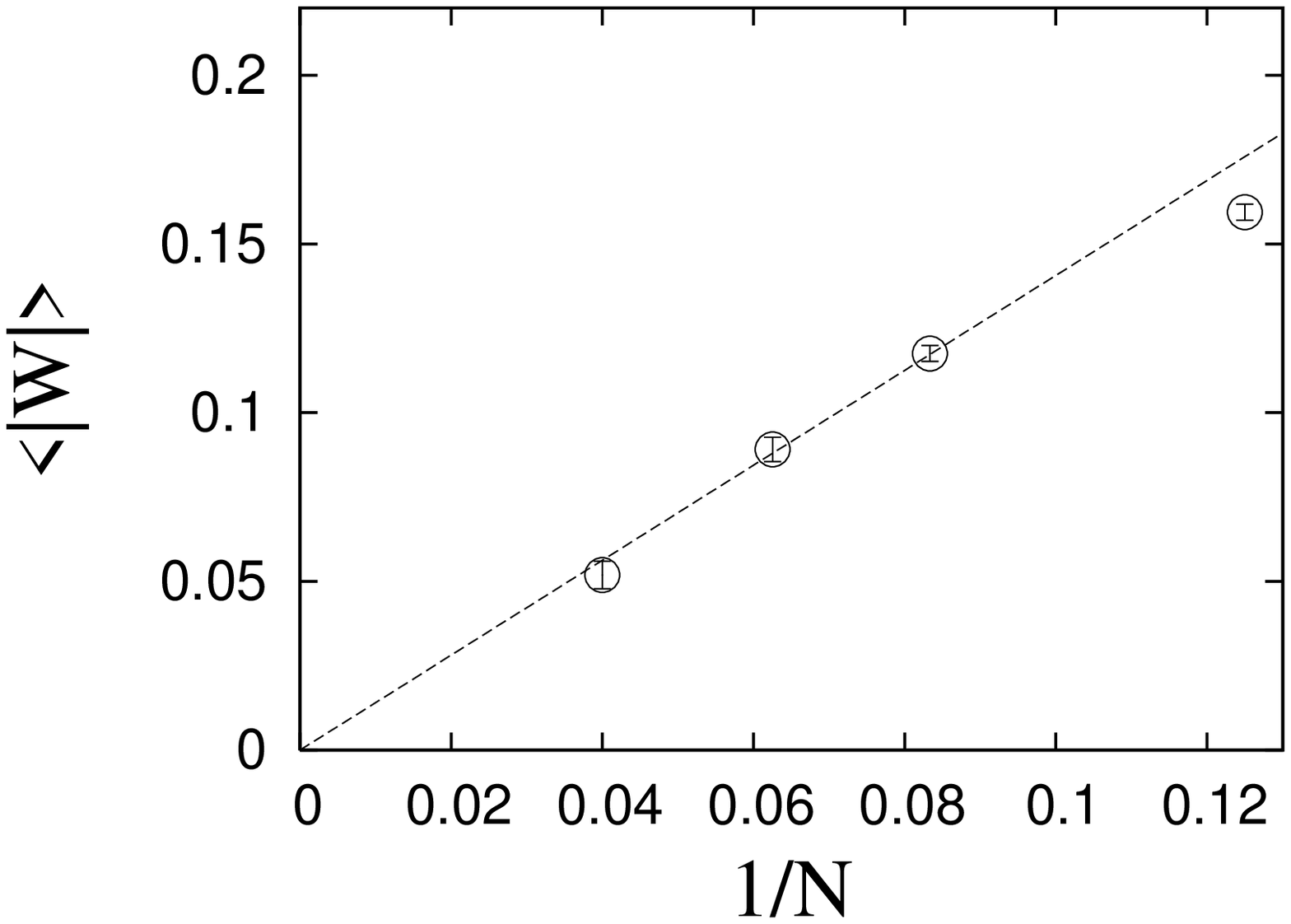}}}
\caption{
$\langle |W|\rangle$ at $b=0.50, \kappa=0.10$, $N_t=1$ for (nontwisted) QCD(Adj). 
}
\label{fig:Wilson_apbc_non_twist}
\end{center}
\end{minipage}
\begin{minipage}{0.33\hsize}
\end{minipage}
\begin{minipage}{0.33\hsize}
\begin{center}
\vspace{-2.4cm}
\rotatebox{-90}{
\scalebox{0.33}{
\includegraphics{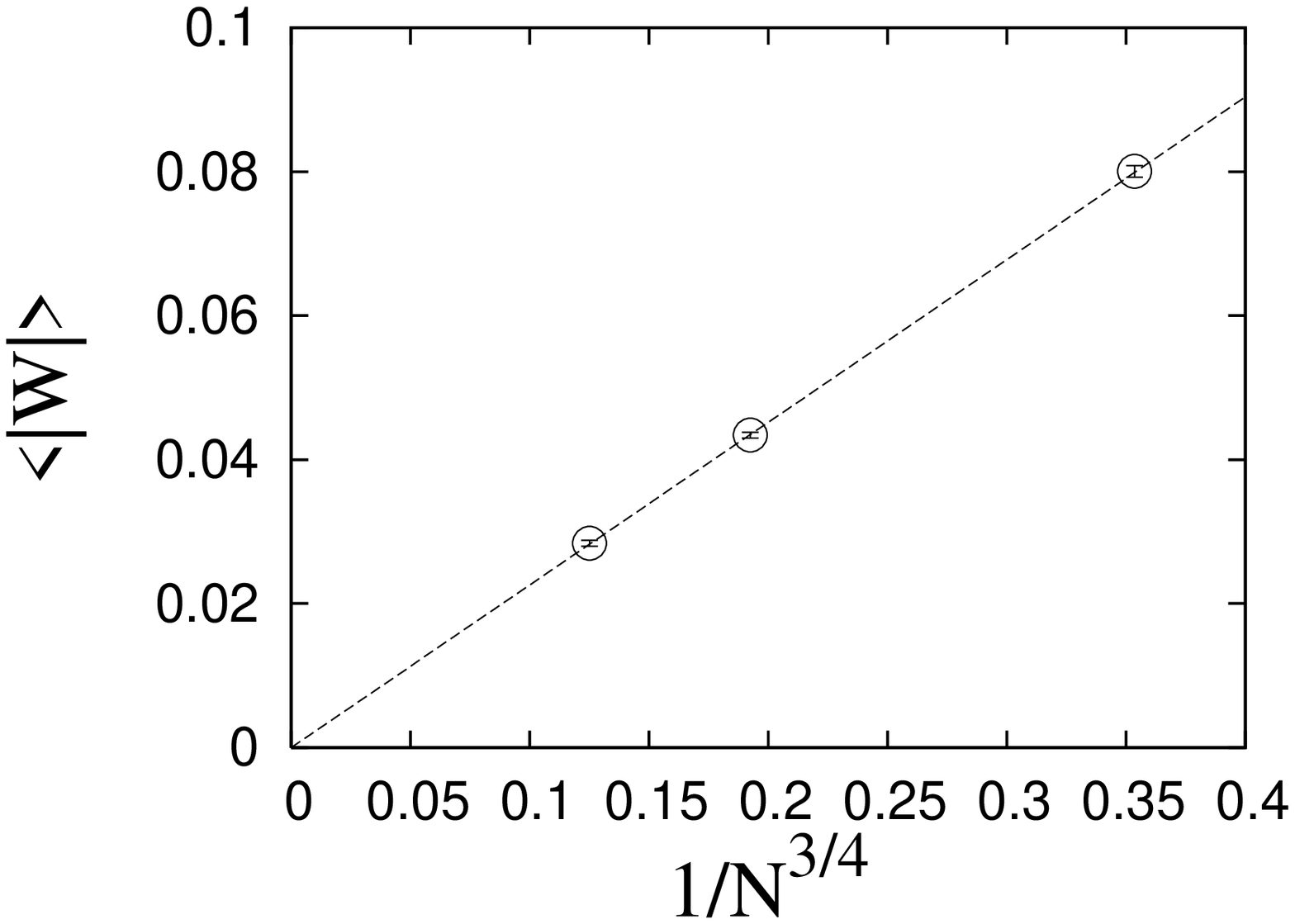}}}
\caption{
$\langle |W|\rangle$ at $b=0.50, \kappa=0.10$, $N_t=1$ for TQCD(Adj). }
\label{fig:Wilson_apbc}
\end{center}
\end{minipage} 
\end{tabular}
\end{figure}

First let us see that  the $(\Z_N)^3$  center symmetry  is not broken. 
In Fig.~\ref{fig:Wilson_apbc_kappa_dependence}, 
we plot the expectation value of an averaged  Wilson line 
given by 
\begin{eqnarray}
|W|\equiv \frac{1}{3N_t}\sum_{i=1}^3\sum_{t=1}^{N_t} |V_{i}(t)|,  
\end{eqnarray}
at $b=0.5$ and various $\kappa$, and $N_t=1$ for QCD(Adj). 
As expected, the $(\Z_N)^3$ is not broken when 
quarks are  sufficiently light for both models. We can see  similar behavior 
for the TQCD(Adj) except for the finite-$N$ corrections.
From Figs.~\ref{fig:Wilson_apbc_non_twist} and \ref{fig:Wilson_apbc}, 
it is clearly seen that $\langle |W|\rangle$ goes to zero as 
$1/N$ for the QCD(Adj) while $1/N^{3/4}$ for the TQCD(Adj).  
That the Wilson line behaves as $\langle|W|\rangle\sim N^{-3/4}=1/\sqrt{V}$ for the TQCD(Adj) 
is the same as what happened in the model in the previous section.  
The same pattern can  also be observed with $N_t>1$ . 

\begin{figure}
\begin{tabular}{ccccc}
\begin{minipage}{0.33\hsize}
\begin{center}
\rotatebox{-90}{
\scalebox{0.33}{
\includegraphics{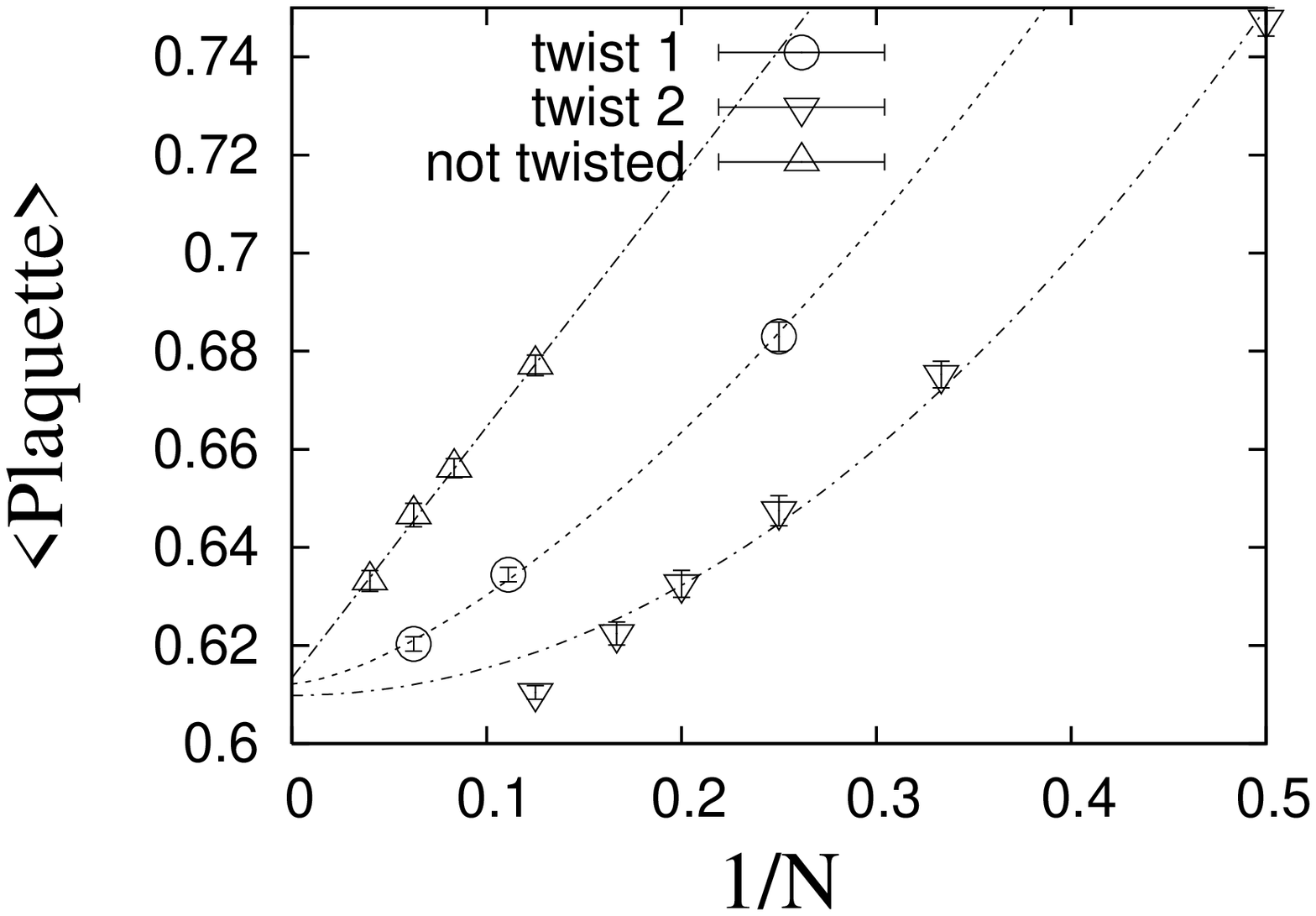}}}
\caption{ 
Expectation value of the plaquette along spatial directions, 
$b=0.50, \kappa=0.10$, $N_t=1$. 
Fitting Ans\"atze  are $a+b/N$ and $a+b/N^{3/2}$ and $a+b/N^2$ for nontwisted, twist 1, and 
twist 2 models, respectively. For twist 2, the $N=8$ point is not used for fitting. 
 }
\label{fig:PlaqS1B050K010}
\end{center}
\end{minipage}
\begin{minipage}{0.33\hsize}
\end{minipage}
\begin{minipage}{0.33\hsize}
\begin{center}
\vspace{-2.4cm}
\rotatebox{-90}{
\scalebox{0.33}{
\includegraphics{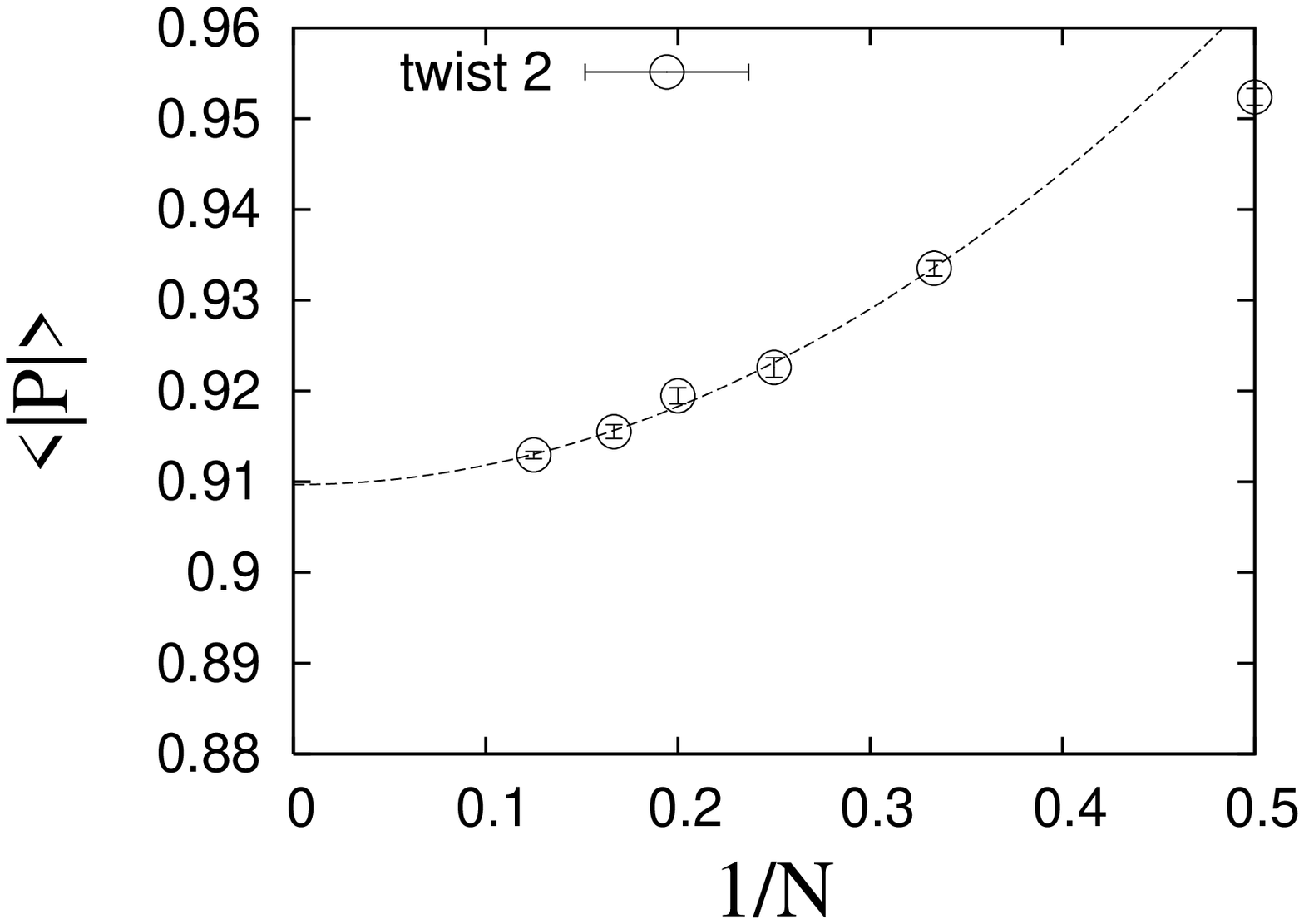}}}
\caption{ 
Expectation value of the Polyakov loop, 
$b=0.50, \kappa=0.10$, $N_t=1$. 
Fitting Ansatz is $a+b/N^2$. 
 }
\label{fig:PlaqS1B050K010new}
\end{center}
\end{minipage}
\begin{minipage}{0.33\hsize}
\end{minipage}
\begin{minipage}{0.33\hsize}
\begin{center}
\vspace{-2.4cm}
\rotatebox{-90}{
\scalebox{0.33}{
\includegraphics{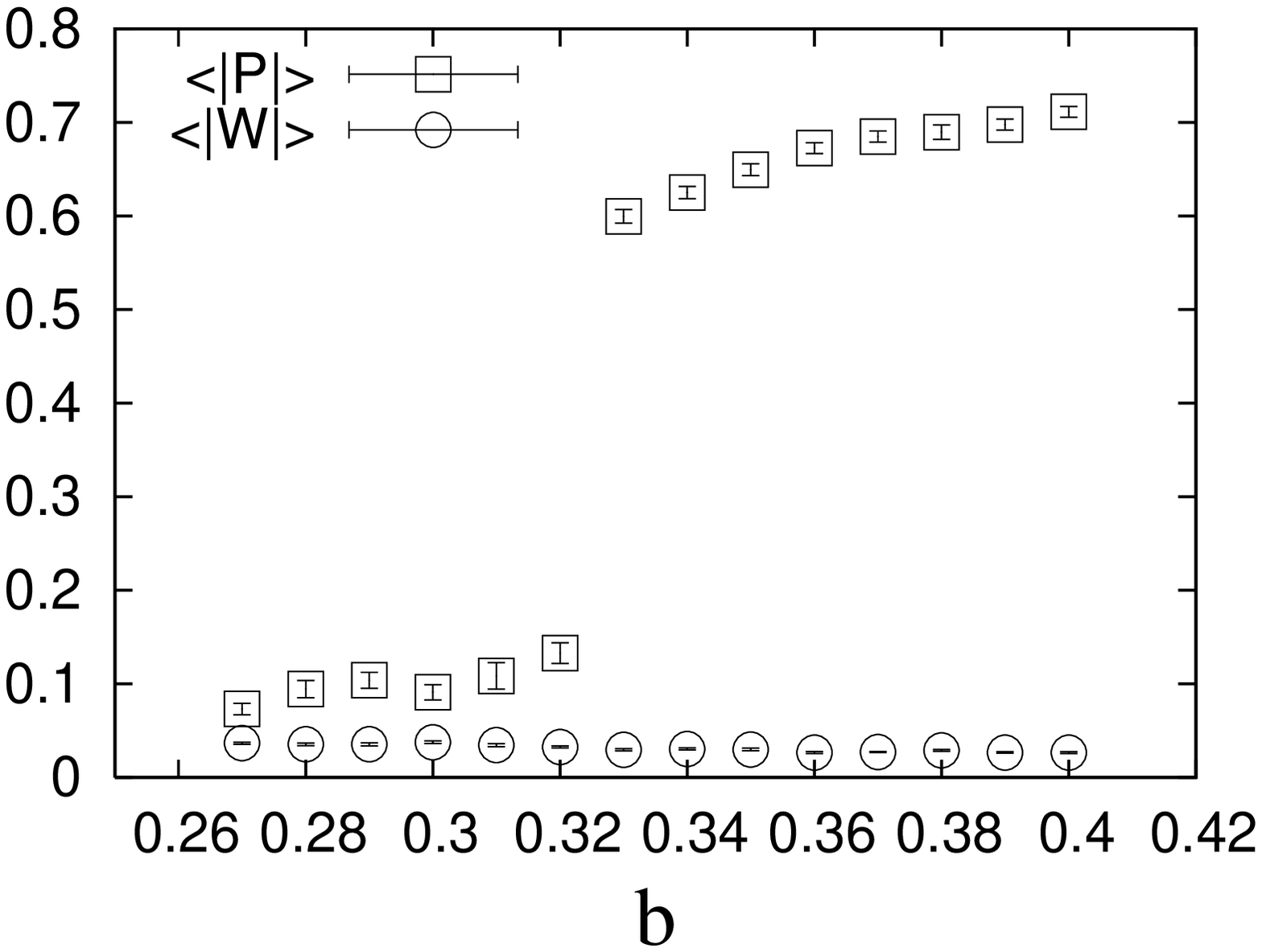}}}
\end{center}
\caption{Expectation values of the Polyakov loop  and Wilson line at $N=16, N_t=2, \kappa=0.10$.}  
\label{deconfinement}
\end{minipage}
\end{tabular}
\end{figure}

In Fig.~\ref{fig:PlaqS1B050K010}, 
the plaquette is  plotted for the QCD(Adj) and TQCD(Adj) with two types of twist. 
The Ansatz $const. + const./N^{3/2}$ for the twist 1  is consistent with the data, 
and by assuming it, the extrapolated value at $N=\infty$ agrees with the one obtained 
from the nontwisted model.  For twist 2,  $const. + const./N^{2}$   is consistent with the data. 
The similar behavior can be seen for the expectation value of the Polyakov loop 
defined by
\begin{eqnarray}
|P| = \frac{1}{N}\left|Tr \displaystyle{\prod_{t=1}^{N_t}}U_t(t) \right|.
\end{eqnarray}
We observe that by using twist 2, we can suppress finite-$N$ corrections  more. 
We notice that, up to $N=25$, we do not observe the jump in 
the expectation value of the plaquette, which corresponds to the bulk transition.

Now,  let us consider the confinement/deconfinement phase transition
in the large-$N$ reduced model at finite temperature. We take $\kappa=0.10$.  
From the  experience in pure bosonic Yang-Mills theory, 
which is studied extensively in \cite{Lucini:2005vg, Panero:2009tv},  
it is known that one has to take the number of spatial links 
$N_s$ to be large (roughly $N_s \gtrsim 2N_t$) 
in order to see the deconfinement  transition clearly. 
For QCD(Adj) without twist  
this is rather severe constraint, because 
$N_s^{\rm eff}$ is related to the number of colors $N$ by $N_s^{\rm eff} \sim N^{1/3}$.  
As a result, at $N_t=2$, we could not observe the transition below $N=30$, 
although we  could observe a clear transition for $N_t=1$. 
In contrast, in TQCD(Adj), the transition can be seen at small values of $N$.  
In Fig.~\ref{deconfinement} we plot the expectation value of the Wilson line  and the Polyakov loop 
for $N_t=2$.
We can clearly see a jump of the expectation values of the Polyakov loop
around $b=0.33$ for $N_t=2$, which corresponds to the confinement/deconfinement 
transition.\footnote{In order to take the  continuum limit, we need to analyze large $N_t$. 
However, to do this, we also need to take $N$ large such that $N_s^{eff}\gtrsim 2N_t$. 
It is beyond our current numerical resources and 
we leave detailed analysis for larger $N_t$ for future work. }   
This result is consistent with the one given by large volume simulations for pure Yang-Mills theory 
 (for example, $SU(8)$ YM theory simulated  on the $ 10^3 \times 5$ lattice  gives, in the extrapolated infinite volume limit,  $b_c \sim 0.34$  \cite{Lucini:2005vg,Panero:2009tv}), 
confirming the finite-temperature version of equivalence nonperturbatively.

Before closing this section, let us  give an estimate for the the physical temperature of the 
transition. By using the 2-loop beta function of 
the bosonic Yang-Mills theory (neglecting fermions because they are heavy), 
$b$ is related to be the temperature $T$ and lattice lambda parameter $\Lambda_{LAT}$ as
\begin{eqnarray}\label{twoloop}
\frac{T}{\Lambda_{LAT}}
=
\frac{1}{N_t}
\left(
\frac{11}{24\pi^2}\lambda
\right)^{51/121}
\exp\left(
\frac{12\pi^2}{11\lambda}
\right),  
\end{eqnarray}
where $\lambda=1/(2b)$. If we take $\Lambda_{LAT}$ to be of order $\mathcal{O}(1$MeV$)$ 
as in the QCD with $SU(3)$, by substituting it to \eqref{twoloop}, 
the transition temperature turns out to be of order $\mathcal{O}(100$MeV$)$, 
as expected.  
\section{Stabilizing  noncommutative Yang-Mills theory}
\hspace{0.51cm}
There is a well-know perturbative  equivalence  between the TEK model and noncommutative YM theory. 
 Nonperturbatively, this  relation is problematic due to a global  instability 
\cite{BNSV06,TV06,AHHI07,AHH08}.  In the perturbative description, the twist-eater background is used to generate a fuzzy torus, the 
  noncommutative base space of target theory. The spontaneous center-symmetry breaking in the TEK model is associated with the spontaneous collapse 
  of the noncommutative torus. 
This instability, in effect, is  related to the center-symmetry breaking instability of the TEK model. 
 For a nice discussion of the relation  between noncommutative theories and matrix models, see, e.g., \cite{AIIKKT99}.

In this work, we suggest that  TQCD(Adj) 
can be used to define  Yang-Mills theory on non-commutative space (for recent developments,  
see e.g. \cite{Steinacker10}). 

As asserted above, TEK construction is problematic due to instability of  the twist-eater background \cite{AHHI07,AHH08,BNSV06,TV06}. 
 Similar construction with the fuzzy sphere 
also fails \cite{AHH08}. 
In noncommutative field theory, this instability 
corresponds to tachyonic modes in gluon propagator \cite{VanRaamsdonk01,AL01}. 

When adjoint fermions are introduced, the situation is different.  
The center symmetry is stabilized even at the one-site lattice and the twist-eater configuration is stable.  
Hence TQCD(Adj) with appropriate parameter scalings can
serve as a nonperturbative formulation of noncommutative Yang-Mills theory with 
adjoint fermions. The  interesting point is that the  background is stable 
even if fermions are very heavy. In such a situation, low energy physics is almost 
a bosonic one, but instability  in the one-site model  is removed thanks to 
fermion induced stabilizing effects. We expect  that the same effect can also be achieved 
by  introducing   a twist to the double-trace deformation of  the Eguchi-Kawai model 
\cite{UY08}, which is purely bosonic.  

\section{Conclusions and Discussions}
\hspace{0.51cm}

In this paper, we investigated the volume reduction for large-$N$ gauge 
theory with adjoint fermions \cite{KUY07}. We used perturbative one-loop analysis crucially supplemented with the estimates of nonperturbative quantum fluctuations to analytically 
explain the zero temperature result of  Ref. \cite{BS09}. We have shown that for heavy fermions 
there exists a large-small volume equivalence, with an intermediate volume dependent phase. 
In the sense of dynamics, the theory exhibits a 4d-0d-4d cascade of dimensions. We have used the small volume phase to extract results for large volume theory. 

 Next, we  generalized 
the volume independence to finite temperature, and confirmed it numerically. In effect, 
 we have constructed a 
one-dimensional lattice model and analyzed it by using Monte Carlo simulation 
to directly show that the center symmetry is preserved for a wide parameter 
region including  heavy fermions. In particular, we introduced a twisted 
version of the large-$N$ reduced theory  with adjoint fermions for numerical 
efficiency and then succeeded in observing 
the confinement/deconfinement transition. The temperature agrees with large volume simulations of pure Yang-Mills theory.
 We also argued that TQCD(Adj) with appropriate 
parameter scaling can serve as a nonperturbative formulation of 
noncommutative Yang-Mills theory.

There are several directions for future research. 
Our discussion of   QCD(Adj) with the $N_f^D=1$  fermion can be generalized to address mass spectrum and other nonperturbative aspects. 
By introducing more fermion flavors, 
 our reduced models may be used to address the  conformality or confinement problem, to address 
 the determination of the lower boundary of the conformal window and  perhaps to study models relevant to the technicolor scenario. 

In this work, we have seen that due to nonperturbative quantum fluctuations, 
the restoration 
of full center symmetry occurs not at $mLN \sim \rm few$, but $mL \sim {\rm few}.$    
This suggests  that the number of double-trace operators suggested in \cite{UY08}  for theories on $T^4$ and  $T^3 \times \R$  is a conservative overestimation, and  might be reduced considerably without spoiling unbroken center symmetry.   
 If so, since deformed reduced  theories are purely bosonic, the simulation cost becomes cheaper and  it can be a practical tool. If it works, it can  be used as a template to study QCD with fundamental matter.

\section*{Acknowledgements}
\hspace{0.51cm}
The authors would like to thank Adi Armoni, Barak Bringoltz, Masafumi Fukuma, Antonio Gonzalez-Arroyo,    Hikaru Kawai, Erich Poppitz,  Steve Sharpe, and  Larry Yaffe for stimulating discussions and comments. 
Very special thanks to  Barak Bringoltz for providing his numerical data,  
which was very useful for checking consistency of our simulation code.  
The authors would also like to thank organizers and participants of 
``MCFP Workshop on Large-$N$ Gauge Theories" for hospitality and useful conversations.   
T.A. would like to thank   
Weizmann Institute of Science and Albert Einstein Institute for hospitality, 
where a part of this work was done; he is supported by the Japan Society for 
the Promotion of Science (JSPS) and by the grant-in-aid for the Global COE program ``The Next Generation of Physics, Spun from Universality
and Emergence''  from the MEXT.
M.\"U. expresses his gratitude  to 
Weizmann Institute of Science for hospitality; his work was supported by
 the U.S.\ Department of Energy Grant DE-AC02-76SF00515. 
 M.H.  and  M.\"U.  thank Aspen Center for Physics where portions of this work was done.  
The numerical computations in this work were in part carried out
on clusters at the Yukawa Institute.   

\appendix

\section{ QCD(Adj)  on continuum $T^4$ at finite-$L$ and  one-loop analysis  } 
\label{Poisson}

In this appendix, we will rewrite the one-loop potential  on small-$T^4$    (\ref{smallT4}) in terms of Wilson lines by using   Poisson resummation.  Poisson   resummation is  a duality which maps a sum over 
KK-momenta given in    (\ref{smallT4})    to a sum over a winding number.  
The two are equivalent expressions, and some aspect of physics is more transparent in one of the two.

Since  (\ref{smallT4}) is periodic under $\theta_{\mu}^{ab} \rightarrow \theta_{\mu}^{ab} + 2\pi$, it can  be Fourier expanded:   
\begin{equation}
S_{\rm 1-loop}[\theta_{\mu}^{ab}] = \sum_{a<b} \sum_{ \vec n \in \Z^4 \setminus \{{\bf  0}\} }
  e^{i \vec \theta^{ab}\cdot \vec n} P_{\vec n}(mL). 
\end{equation} 
For a given winding number $\vec n\equiv (n_1, \ldots, n_4)$,  
\begin{equation}
\sum_{a < b}   e^{i \vec \theta^{ab}\cdot \vec n}  =  \half \sum_{a ,  b}  e^{i \vec \theta^{ab}\cdot \vec n}   
 -  \half \sum_{a}  \; 1 = \half \left( \left|\tr\left(V_1^{n_1}    \cdots V_4^{n_4} \right)\right|^2 -N\right). 
\end{equation}

It is also useful to express this sum in terms of trace over adjoint representation matrices, 
given by 
 \begin{eqnarray}
 \Omega_{\rm adj}(\vec n) &&= \left( V_1^{n_1} \ldots V_4^{n_4} \right) \otimes  \left( V_1^{n_1} \ldots V_4^{n_4} \right)^{\dagger},   
\end{eqnarray}
or equivalently,
\begin{eqnarray}
 \Omega_{\rm adj}(\vec n) &&  =   \left[ \begin{array}{c|cccc}
{\bf 1}_{N\times N} & &&& \cr
\hline  
& e^{i (\vec \theta_1 - \vec \theta_2)\cdot\vec n} & &&  \cr
&& \ddots & & \cr 
&& & e^{i (\vec \theta_{a} - \vec \theta_{b})\cdot\vec n} & \cr
&& && \ddots
   \end{array} 
  \right].
\end{eqnarray}
Clearly, $\tr  \Omega_{\rm adj} (\vec n)=  \left| \tr \left(V_1^{n_1}    \cdots V_4^{n_4} \right)\right|^2$.  
 Equation (\ref{smallT4}) can dually be  rewritten as a sum over winding modes
 \begin{eqnarray}
    S_{\rm 1-loop}[V_1, \ldots,  V_4 ] &&= 
    \frac{2}{\pi^2} \sum_{a<b}
    \sum_{ \vec n \in \Z^4 \setminus \{{\bf  0}\}}  \;   
     \frac{1}{ |\vec n|^4} \left(- 1+  N_f^D m^2 L^2  |\vec n|^2  K_2(  mL  |\vec n| ) \right)
    \cdot\cos( \vec n\cdot \vec  \theta^{ab} ) \cr 
    && \equiv  
     \frac{1}{\pi^2}  
         \sum_{ \vec n \in \Z^4 \setminus \{{\bf  0}\}}  \;   m_{\vec n}^2 \;
          \left( \left|\tr\left(V_1^{n_1}    \cdots V_4^{n_4} \right)\right|^2 -N\right),  
 \label{MMP}
\end{eqnarray}
where  $m_{\vec n}^2$ is interpreted as the mass square  of the Wilson line with winding number $\vec n$.     
Unlike the similar  sums appearing in  one-loop effective potential  on $\R^{4-d} \times T^{d}$ with $1 \leq d \leq 3$ which  are  absolutely convergent \cite{KUY07},  the series (\ref{MMP}) is 
{\it conditionally convergent}.\footnote{In (\ref{MMP}), the subtraction of  the constant divergent term is related 
to the absence of log divergent and holonomy independent $\sum_{a=b}$ terms in the first line, as well as (\ref{smallT4}). 
With this subtraction, any singularity that may appear in the sum    (\ref{MMP}) or its dual (\ref{smallT4}) is physical IR-singularity. 
  IR-aspects of one-loop effective action can be examined  in either picture.   }
  This is, of course, physical and related to   nontrivial infrared (IR) aspects of the theory 
  which we discuss below.  

\subsection{IR singularities, conditional convergence, and noncommutative saddles} 
The series in (\ref{smallT4}) and (\ref{MMP}) are equivalent expressions, related to each other via Poisson resummation.  Consider the zero  KK-momenta subsector of (\ref{smallT4}).   Since eigenvalue separation has an interpretation as momentum,  (\ref{smallT4})
   exhibits an IR singularity whenever   $\sum_{\mu=1}^{4} (\theta_{\mu}^{ab})^2=0$ for some $a, b$. This IR problem is also manifest in lattice one-site one-loop action  (\ref{1loop-l}).  The series (\ref{MMP}) is conditionally convergent, and whenever two eigenvalues are coincident,  
$   \sum_{ \vec n \in \Z^4 \setminus \{{\bf  0}\}} $ exhibits logarithmic  IR divergences due to modes with large-winding number $|\vec n| \rightarrow \infty$.  Winding modes have an interpretation in terms of spacetime distance  \cite{Bringoltz09} and this is the same IR problem  as in (\ref{smallT4}). The physical interpretation of divergence is  as follows:  Whenever two (or more) eigenvalues are coincident,   there are (in perturbation theory) massless modes (analog of $W$ bosons or open strings). IR divergence  comes about because we have integrated out  these massless modes that we should have kept  in the correct description of long-distance dynamics of the effective theory. 

Whenever two (or more)  $\sum_{\mu=1}^{4} (\theta_{\mu}^{ab})^2=0$, the zeroth order assumption that one can expand the fluctuations around commuting saddles  (\ref{comsad})  is incorrect . There are circumstances where    commutative saddle points of the classical theory 
may  be a good description at one-loop order in perturbation theory for some range of $m$.  For example, if $m=0$, then, the  one-loop effective action  (\ref{smallT4}) and (\ref{MMP}) 
reduces to 
  \begin{eqnarray}
  S_{\rm 1-loop}[\theta_{\mu}^{ab}] &&=   (2-4N_f^D) \sum_{a<b}  \sum_{k_1, \ldots, k_4}    
    \;  \log 
\left( \sum_{\mu=1}^4  \frac{(2 \pi k_{\mu} + \theta_{\mu}^{ab})^2}{L^{2}} \right)  \cr
     &&=
   (-1+ 2N_f^D)   \frac{1}{\pi^2}
    \sum_{ \vec n \in \Z^4 \setminus \{{\bf  0}\}} \frac{1}{ |\vec n|^4} 
     \left( \tr  \Omega_{\rm adj} (\vec n)
      -N \right) .
\label{MMP2}
\end{eqnarray}
This  one-loop action  is bounded from below  and generates a repulsion between eigenvalues.

If $m=\infty$ is sufficiently large, then  the  one-loop action   (\ref{smallT4}) and (\ref{MMP}) is unbounded 
from below and leads to an attraction between eigenvalues. 
 A configuration where  all eigenvalues clump and center is broken is  the minimum. However,  the action  is IR divergent there.  This is an artifact of perturbation theory. Whenever  eigenvalues are close to  each other, then the relevant scale in the theory is  $\lambda_{0d}^{1/4}$, and 
the  dynamics is strongly coupled  for eigenvalues in the $  |\vec x_{ab}|  \lesssim \lambda_{0d}^{1/4}  $ domain where one should not use perturbation theory. 

At one-loop order in perturbation theory, the  actions  (\ref{smallT4}) and (\ref{MMP}) as well as the one-site version  (\ref{1loop-l}) realize all the saddles conjectured to exist in Ref.\cite{Coste:1985mn}.  At $m=0$, the leading fluctuations are   quadratic or Gaussian; at $m=\infty$,  the leading fluctuations are quartic  according to the  classification of Ref.\cite{Coste:1985mn}. As $mL$  is dialed, all interesting saddles with varying number of quadratic and quartic fluctuations appear in massive QCD(Adj).   In  Refs.\cite{Coste:1985mn,BHN82}, only the two extreme cases were shown to exist  in  pure Yang-Mills theory in $d$ dimensions, as $d$ is varied.

Since the one-loop action has  IR singularities, the way to obtain the set of saddle points requires  some care. 
\begin{enumerate} 
\item Introduce an auxiliary   IR cutoff $\mu_{IR} \ll m$, modifying gauge contribution in 
(\ref{smallT4}) to $ \sum_{\mu=1}^4  (2 \pi k_{\mu} +  \theta_{\mu}^{ab})^2L^{-2} + \mu_{IR}^2 $  \cite{BS10}.
 At any finite but infinitesimal value of $\mu_{IR} $, we can sensibly compare the one-loop effective action of  different saddles, and find the   global minima  at a given value of $mL$.  At $m=\infty$, the global minima are at 
$V_{\mu} =1$ (and its center conjugates).   This  can be studied by using the Hermitian matrix model (\ref{MM action}). 

\item  
Assume that the global minimum is a  $k$-bunch  configuration of Wilson line phases. 
The nonperturbative width  of each clump, due to quantum fluctuations is determined by 
zero-dimensional matrix model. 
Interactions between  different clumps are well approximated by one-loop effective action, and are repulsive. 
Dynamics inside each clump is approximated by the $SU(N/k)$ matrix model, 
\begin{eqnarray}
S_{\rm clump}
&=&
\frac{(N/k)}{(\lambda_{0d}/k)}
\ Tr\Biggl(
-
\frac{1}{4}[X^{\prime\mu},X^{\prime\nu}]^2 
+
\sum_{f=1}^{N_f^D}
\bar{\psi}'_f\left(
\gamma_{\mu}[X^{\prime\mu},\psi'_f ]
+
m\psi'_f
\right)
\Biggl), 
\end{eqnarray}
where we put primes in order to emphasize that matrices are $(N/k)\times (N/k)$. 
Then, 't Hooft coupling effectively becomes $\lambda_{0d}/k$, and hence the nonperturbative width of each clump is   $\sim(\lambda_{0d}/k)^{1/4}$ . 

\item 

 At  relatively large values of fermion mass,  $mL \sim 1$, 
 one may expect that only a  small  subgroup of  $({\mathbb Z}_N)^4$ 
 symmetry persists. 
 However, if $k$ is sufficiently  large that distance between bunches becomes smaller than the
 fluctuation scale, $2\pi/(Lk^{1/4})\lesssim\lambda_{0d}^{1/4}=\lambda_{4d}^{1/4}/L$, 
 interaction between nearby bunches cannot be evaluated by one-loop approximation, 
 and distinction of bunches becomes obscure.   
 This implies that the $k$-bunch phase is indistinguishable from the uniform phase. 
  (See Fig.\ref{region5}). 
\begin{figure}[htbp]
\begin{center}
\scalebox{0.2}{
\includegraphics{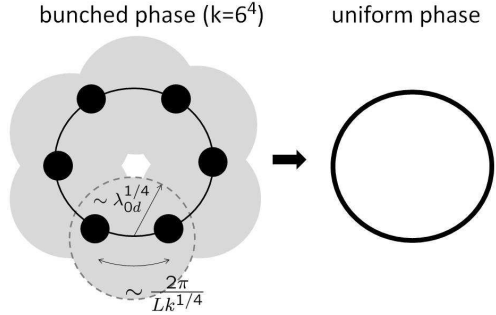}}
\end{center}
\caption{If $k$ is  large enough so that the  distance between two nearest bunches $2\pi/(Lk^{1/4})$ 
becomes smaller than the quantum fluctuation scale $\lambda_{0d}^{1/4}$, interaction between 
these two bunches cannot be evaluated by using one-loop effective action. 
Quantum  fluctuation turns $k$-bunch phase into the uniform phase. } 
\label{region5}
\end{figure}
\end{enumerate}


\end{document}